\documentclass[12pt,a4paper]{article}
\usepackage{graphicx}
\usepackage[T1]{fontenc}
\usepackage[utf8]{inputenc}
\usepackage{textcomp}
\usepackage[sc,osf]{mathpazo}
\usepackage{a4wide}  
\usepackage{latexsym,amsthm,amsfonts,amsmath,mathrsfs,amssymb}
\usepackage{dsfont}
\usepackage{accents}
\usepackage[nosort]{cite}
\usepackage{booktabs} % for much better looking tables
\usepackage[unicode,implicit]{hyperref}
\hypersetup{%
  pdftitle    = {O(n,n) invariance and Wald entropy formula in the Heterotic Superstring effective action at first order in alpha'}
  pdfkeywords = {duality, T duality, strings, heterotic superstrings, entropy,
  black holes, Wald, compactification}
  pdfauthor   = {Tomas Ortin},
  plainpages  = true,
  colorlinks  = true,
  citecolor   = blue,
  urlcolor    = red,
  linkcolor   = black
}
\newcommand{\hepth}[1]{{\tt
\href{http://www.arXiv.org/abs/hep-th/#1}{hep-th/#1}}}
\newcommand{\grqc}[1]{{\tt
\href{http://www.arXiv.org/abs/gr-qc/#1}{gr-qc/#1}}}

\newcommand{\arxiv}[1]{{\tt arXiv:\href{http://www.arXiv.org/abs/#1}{#1}}}
\newcommand{\doi}[1]{{\tt DOI:\href{http://dx.doi.org/#1}{#1}}}

\makeatletter
\@addtoreset{equation}{section}
\makeatother

\begin{document}

\begin{flushright}
\small
IFT-UAM/CSIC-20-073\\
\texttt{arXiv:2005.14618 [hep-th]}\\
May 30\textsuperscript{th}, 2020\\
\normalsize
\end{flushright}

\vspace{1cm}

\begin{center}

  {\Large {\bf O$(n,n)$ invariance and Wald entropy formula\\[.5cm]
      in the Heterotic Superstring effective action\\[.5cm]
      at first order in $\alpha'$}}

\vspace{2.5cm}

\renewcommand{\thefootnote}{\alph{footnote}}

{\sl\large Tom\'{a}s Ort\'{\i}n}\footnote{Email: {\tt tomas.ortin[at]csic.es}}

\setcounter{footnote}{0}
\renewcommand{\thefootnote}{\arabic{footnote}}

\vspace{1cm}

{\it Instituto de F\'{\i}sica Te\'orica UAM/CSIC\\
C/ Nicol\'as Cabrera, 13--15,  C.U.~Cantoblanco, E-28049 Madrid, Spain}

\vspace{2cm}

%%%%%%%%%%%%%%%%%%%%%%%%%%%%%%%%%%%%%%%%%%%%%%%%%%%%%%%%%%%%%%%%%%%%%%

{\bf Abstract}
\end{center}
\begin{quotation}
  {\small We perform the toroidal compactification of the full Bergshoeff-de
    Roo version of the Heterotic Superstring effective action to first order
    in $\alpha'$. The dimensionally-reduced action is given in a
    manifestly-O$(n,n)$-invariant form which we use to derive a
    manifestly-O$(n,n)$-invariant Wald entropy formula which we then use to
    compute the entropy of $\alpha'$-corrected, 4-dimensional, 4-charge,
    static, extremal, supersymmetric black holes.}
\end{quotation}

\newpage
%%%%%%%%%%%%%%%%%%%%%%%%%%%%%%%%%%%%%%%%%%%%%%%%%%%%%%%%%%%%%%%%%%%%%%
%%%%%%%%%%%%%%%%%%%%%%%%%%%%%%%%%%%%%%%%%%%%%%%%%%%%%%%%%%%%%%%%%%%%%%
%%%%%%%%%%%%%%%%%%%%%%%%%%%%%%%%%%%%%%%%%%%%%%%%%%%%%%%%%%%%%%%%%%%%%%
%%%%%%%%%%%%%%%%%%%%%%%%%%%%%%%%%%%%%%%%%%%%%%%%%%%%%%%%%%%%%%%%%%%%%%
\pagestyle{plain}
%%%%%%%%%%%%%%%%%%%%%%%%%%%%%%%%%%%%%%%%%%%%%%%%%%%%%%%%%%%%%%%%%%%%%%
%%%%%%%%%%%%%%%%%%%%%%%%%%%%%%%%%%%%%%%%%%%%%%%%%%%%%%%%%%%%%%%%%%%%%%
%%%%%%%%%%%%%%%%%%%%%%%%%%%%%%%%%%%%%%%%%%%%%%%%%%%%%%%%%%%%%%%%%%%%%%
%%%%%%%%%%%%%%%%%%%%%%%%%%%%%%%%%%%%%%%%%%%%%%%%%%%%%%%%%%%%%%%%%%%%%%

\tableofcontents

%\newpage

%%%%%%%%%%%%%%%%%%%%%%%%%%%%%%%%%%%%%%%%%%%%%%%%%%%%%%%%%%%%%%%%%%%%%%
%%%%%%%%%%%%%%%%%%%%%%%%%%%%%%%%%%%%%%%%%%%%%%%%%%%%%%%%%%%%%%%%%%%%%%
%%%%%%%%%%%%%%%%%%%%%%%%%%%%%%%%%%%%%%%%%%%%%%%%%%%%%%%%%%%%%%%%%%%%%%
%%%%%%%%%%%%%%%%%%%%%%%%%%%%%%%%%%%%%%%%%%%%%%%%%%%%%%%%%%%%%%%%%%%%%%
\section*{Introduction}
%%%%%%%%%%%%%%%%%%%%%%%%%%%%%%%%%%%%%%%%%%%%%%%%%%%%%%%%%%%%%%%%%%%%%%
%%%%%%%%%%%%%%%%%%%%%%%%%%%%%%%%%%%%%%%%%%%%%%%%%%%%%%%%%%%%%%%%%%%%%%
%%%%%%%%%%%%%%%%%%%%%%%%%%%%%%%%%%%%%%%%%%%%%%%%%%%%%%%%%%%%%%%%%%%%%%
%%%%%%%%%%%%%%%%%%%%%%%%%%%%%%%%%%%%%%%%%%%%%%%%%%%%%%%%%%%%%%%%%%%%%%

In a recent paper \cite{Elgood:2020xwu} we have performed the dimensional
reduction of the Heterotic Superstring effective action in a circle to
first-order in $\alpha'$ with two goals in mind:

\begin{enumerate}
\item To study T~duality in the dimensionally-reduced theory and the effect
  that the first-order $\alpha'$ corrections have in it. In particular, we
  wanted to recover the first-order in $\alpha'$ corrections to the Buscher
  T~duality rules \cite{Buscher:1987sk,Buscher:1987qj} found in
  Ref.~\cite{Bergshoeff:1995cg},\footnote{See also
    \cite{Serone:2005ge,Bedoya:2014pma}.} and show explicitly that the whole
  action is invariant under them to that order.

\item To derive a T~duality-invariant formula for the Wald entropy using the
  Iyer-Wald prescription developed in
  Refs.~\cite{Lee:1990nz,Wald:1993nt,Iyer:1994ys}.\footnote{A discussion of
    the caveats in the direct use of this prescription in the Heterotic
    Superstring effective action can be found in the Introduction of
    Ref.~\cite{Elgood:2020xwu}. On the other hand, it is clear that the
    entropy formula derived in that reference by using this prescription gives
    results which coincide with those obtained by microstate counting
    \cite{Kraus:2005zm} and also satisfy the fundamental thermodynamic
    relation $\frac{\partial S}{\partial M}=\frac{1}{T}$ in black hole with
    finite temperature \cite{Cano:2019ycn}. See also the discussion in the
    Conclusions section.}

\end{enumerate}

This formula, though, can only be applied to black holes which can be obtained
from a solution of the 10-dimensional theory by 1 non-trivial and several
trivial dimensional reductions over circles. This severely limits its
applicability to 5-dimensional black holes and certain 4-dimensional ones.

It is natural to try to extend those results to non-trivial toroidal
compactifications, testing the O$(n,n)$ invariance of the
dimensionally-reduced action\footnote{If the $n_{V}$ 10-dimensional gauge
  fields are Abelian, the theory is expected to have a larger duality group:
  O$(n,n+n_{V})$ \cite{Maharana:1992my}, but this group is obviously broken
  when they are non-Abelian, since they cannot be rotated into the Abelian
  Kaluza-Klein and winding vector fields. Here we will focus mostly on the
  O$(n,n)$ duality group which is expected to always be present.} to first
order in $\alpha'$ and obtaining a manifestly O$(n,n)$-invariant Wald entropy
formula that can be applied to more general black-hole solutions such as, for
instance, the heterotic version of the 4-dimensional, 4-charge, static,
extremal black holes whose microscopic entropy was first computed in
Refs.~\cite{Maldacena:1996gb,Johnson:1996ga}.

Earlier work on the effect of $\alpha'$ corrections on the T~duality
invariance of the Heterotic Superstring effective action more or less complete
in different forms and schemes \cite{Metsaev:1987zx}, including Double Field
Theory, can be found in
Refs.~\cite{Meissner:1996sa,Kaloper:1997ux,Hohm:2014eba,Marques:2015vua,Baron:2017dvb,Eloy:2019hnl,Eloy:2020dko}
some of which we will comment upon in the main body of this paper. Here we
will use the Bergshoeff-de Roo action Ref.~\cite{Bergshoeff:1989de} obtained
by supersymmetric completion of the Lorentz Chern-Simons terms in the
Kalb-Ramond field strength \cite{Bergshoeff:1988nn}.

This paper is organized as follows: in Section~\ref{sec-heteroticalpha} we
introduce the 10-dimensional Heterotic Superstring effective action in the
Bergshoeff-de Roo formulation. In Section~\ref{sec-dimredO1}, as a warm-up
exercise, we review the toroidal dimensional reduction of the zeroth-order
action, rewriting it in a manifestly O$(n,n)$-invariant form. In
Section~\ref{sec-reductionheteroticsupergravity} we add Yang-Mills fields, and
rewrite the dimensionally-reduced action in an apparently manifestly
O$(n,n+n_{V})$-invariant form, reproducing, in the Abelian case (when that
invariance is real), the results of Maharana and Schwarz
\cite{Maharana:1992my}. In Section~\ref{sec-completereductiontofirstorder} we
consider the full $\mathcal{O}(\alpha')$ action, which amounts to the addition
of the torsionful spin-connection terms. The full action can be
regarded,formally, as that of the previous section with more gauge fields and
a gauge group which is the direct product of the Yang-Mills gauge group and
the 10-dimensional Lorentz group SO$(1,9)$ \cite{Bergshoeff:1988nn} and, in a
first stage (Section~\ref{sec-firststep}), we can simply use the results of
the previous section. This cannot be the final result, though, because, as
different from the Yang-Mills group, the 10-dimensional Lorentz group is
broken into the $(10-n)$-dimensional one and O$(n)$. Thus, in a second stage
(Section~\ref{sec-secondstep}), we perform this decomposition leaving the
dimensionally-reduced action in a manifestly gauge-, $(10-n)$-dimensional
Lorentz-, diffeomorphism- and O$(n,n)$-invariant form. Then, in
Section~\ref{sec-entropyformula} we derive from that action a Wald entropy
formula that we test on 4-dimensional 4-charge black holes. We present our
conclusions in Section~\ref{sec-conclusions}. Appendix~\ref{app-Onn} contains
relevant formulae concerning the
$\mathrm{O}(n,n)/(\mathrm{O}(n)\times \mathrm{O}(n))$ coset space
that we use in the manifestly-O$(n,n)$-invariant action. 

%%%%%%%%%%%%%%%%%%%%%%%%%%%%%%%%%%%%%%%%%%%%%%%%%%%%%%%%%%%%%%%%%%%%%%
%%%%%%%%%%%%%%%%%%%%%%%%%%%%%%%%%%%%%%%%%%%%%%%%%%%%%%%%%%%%%%%%%%%%%%
%%%%%%%%%%%%%%%%%%%%%%%%%%%%%%%%%%%%%%%%%%%%%%%%%%%%%%%%%%%%%%%%%%%%%%
%%%%%%%%%%%%%%%%%%%%%%%%%%%%%%%%%%%%%%%%%%%%%%%%%%%%%%%%%%%%%%%%%%%%%%
\section{The Heterotic Superstring effective action to
  \texorpdfstring{$\mathcal{O}(\alpha')$}{O(α')}}
\label{sec-heteroticalpha}
%%%%%%%%%%%%%%%%%%%%%%%%%%%%%%%%%%%%%%%%%%%%%%%%%%%%%%%%%%%%%%%%%%%%%%
%%%%%%%%%%%%%%%%%%%%%%%%%%%%%%%%%%%%%%%%%%%%%%%%%%%%%%%%%%%%%%%%%%%%%%
%%%%%%%%%%%%%%%%%%%%%%%%%%%%%%%%%%%%%%%%%%%%%%%%%%%%%%%%%%%%%%%%%%%%%%
%%%%%%%%%%%%%%%%%%%%%%%%%%%%%%%%%%%%%%%%%%%%%%%%%%%%%%%%%%%%%%%%%%%%%%

Let us first introduce the Heterotic Superstring effective action to
$\mathcal{O}(\alpha')$, where $\alpha'$ is the Regge slope
parameter,\footnote{The Regge slope parameter is related to the string length
  $\ell_{s}$ by $\alpha'=\ell_{s}^{2}$.} in the formulation of
Ref.~\cite{Bergshoeff:1989de} but using the conventions of
Ref.~\cite{Ortin:2015hya}.\footnote{The relation between the normalizations of
  the fields in Ref.~\cite{Bergshoeff:1989de} and here can be found in
  Ref.~\cite{Fontanella:2019avn}.}

The torsionful spin connection and Kalb-Ramond field strength, which are two
fundamental ingredients of the action, can be constructed recursively order by
order in $\alpha'$. At zeroth-order, the field strength of the
Kalb-Ramond 2-form $B_{\mu\nu}$ is defined as

\begin{equation}
H^{(0)}{}_{\mu\nu\rho} \equiv 3\partial_{[\mu}B_{\nu\rho]}\, ,
\end{equation}

\noindent
and it is added as torsion to the (torsionless, metric-compatible) Levi-Civita
spin connection 1-form ${\omega}_{\mu}{}^{{a}}{}_{{b}}$ as

\begin{equation}
{\Omega}^{(0)}_{(\pm)\, \mu}{}^{{a}}{}_{{b}} 
=
{\omega}_{\mu}{}^{{a}}{}_{{b}}
\pm
\tfrac{1}{2}{H}^{(0)}{}_{{\mu}}{}^{{a}}{}_{{b}}\, ,
\end{equation}

\noindent
to construct the zeroth-order torsionful spin connections.

The corresponding zeroth-order Lorentz curvature 2-forms and
Chern-Simons 3-forms are defined as

\begin{eqnarray}
  \label{eq:R0def}
{R}^{(0)}_{(\pm)\, \mu\nu}{}^{{a}}{}_{{b}}
& = & 
2\partial_{[\mu|} {\Omega}^{(0)}_{(\pm)\, |\nu]}{}^{{a}}{}_{{b}}
-2 {\Omega}^{(0)}_{(\pm)\, [\mu|}{}^{{a}}{}_{{c}}\,
{\Omega}^{(0)}_{(\pm)\, |\nu]}{}^{{c}}{}_{{b}}\, ,
\\
  & & \nonumber \\
  \label{eq:oL0def}
{\omega}^{{\rm L}\, (0)}_{(\pm)}
& = &  
3{R}^{ (0)}_{(\pm)\, [\mu\nu|}{}^{{a}}{}_{{b}} 
{\Omega}^{ (0)}_{(\pm)\, |\rho]}{}^{{b}}{}_{{a}} 
+2
{\Omega}^{ (0)}_{(\pm)\, [\mu|}{}^{{a}}{}_{{b}} \,
{\Omega}^{ (0)}_{(\pm)\, |\nu|}{}^{{b}}{}_{{c}} \,
{\Omega}^{ (0)}_{(\pm)\, |\rho]}{}^{{c}}{}_{{a}}\, .  
\end{eqnarray}

At first order in $\alpha'$ we also have to take into account the Yang-Mills
fields.  The gauge field is denoted by $A^{A}{}_{\mu}$, where $A,B,C,\ldots$
are the adjoint gauge indices of some group that we will not specify. The
corresponding gauge field strength and the Chern-Simons 3-forms are defined by

\begin{eqnarray}
  \label{eq:FAdef}
{F}^{A}{}_{\mu\nu}
& = & 
2\partial_{[\mu}{A}^{A}{}_{\nu]}+f_{BC}{}^{A}{A}^{B}{}_{[\mu}{A}^{C}{}_{\nu]}\, , 
\\
  & & \nonumber \\
  \label{eq:oYMdef}
{\omega}^{\rm YM}
& = & 
3F_{A\, [\mu\nu}{A}^{A}{}_{\rho]}
-f_{ABC}{A}^{A}{}_{[\mu}{A}^{B}{}_{\nu}{A}^{C}{}_{\rho]}\, ,
\end{eqnarray}

\noindent
where we have lowered the adjoint group indices using the Killing metric of
$K_{AB}$: $f_{ABC}\equiv f_{AB}{}^{D}K_{DB}$ and of the gauge fields
$F_{A\, \mu\nu}\equiv K_{AB}F^{B}{}_{\mu\nu}$.

Then, the first-order Kalb-Ramond field strength is given by 

\begin{equation}
H^{(1)}{}_{\mu\nu\rho}
= 
3\partial_{[\mu}B_{\nu\rho]}
+\frac{\alpha'}{4}\left({\omega}^{\rm YM}{}_{\mu\nu\rho}
+{\omega}^{{\rm L}\, (0)}_{(-)\, \mu\nu\rho}\right)\, ,  
\end{equation}

\noindent
and now it is the torsion of the first-order torsionful spin connection

\begin{eqnarray}
{\Omega}^{(1)}_{(\pm)\, \mu}{}^{{a}}{}_{{b}} 
& = & 
{\omega}_{\mu}{}^{{a}}{}_{{b}}
\pm
\tfrac{1}{2}{H}^{(1)}_{{\mu}}{}^{{a}}{}_{{b}}\, ,
\end{eqnarray}

\noindent
whose curvature ${R}^{(1)}_{(\pm)\, \mu\nu}{}^{{a}}{}_{{b}}$ and Chern-Simons
form ${\omega}^{{\rm L}\, (1)}_{(\pm)}$ are now used to define the
second-order Kalb-Ramond field strength $H^{(2)}$ and so on.

Only
$\Omega^{(0)}_{(\pm)\, \mu},{R}^{(0)}_{(\pm)\, \mu\nu}{}^{a}{}_{b},
\omega^{{\rm L}\, (0)}_{(\pm)\, \mu\nu\rho}$ and $ H^{(1)}{}_{\mu\nu\rho}$
(plus the Yang-Mills fields) occur in the $\mathcal{O}(\alpha')$ action and,
in terms of these objects plus the dilaton field $\phi$ and the Ricci scalar
$R$ of the metric $g_{\mu\nu}$, the first-order in $\alpha'$ Heterotic
Superstring effective action in the string frame takes the form

\begin{equation}
\label{heterotic}
S
=
\frac{g_{s}^{2}}{16\pi G_{N}^{(10)}}
\int d^{10}x\sqrt{|{g}|}\, 
e^{-2\phi}\, 
\left\{
{R} 
-4(\partial\phi)^{2}
+\tfrac{1}{12}H^{2}
-\dfrac{\alpha'}{8}\left[
F_{A}\cdot F^{A}
+
R_{(-)}{}^{a}{}_{b}\cdot R_{(-)}{}^{b}{}_{a}
\right]
\right\}\, ,
\end{equation}

\noindent
here $g_{s}$ is the Heterotic Superstring coupling constant, which is given by
the vacuum expectation value of $e^{\phi}$, the dot indicates the contraction
of the indices of 2-forms:
${F}_{A}\cdot {F}^{A}\equiv {F}_{A\, \mu\nu}{F}^{A\, \mu\nu}$ and the
10-dimensional Newton constant $G_{N}^{(10)}$ is related to the string length
and coupling constants by

\begin{equation}
\label{eq:d10newtonconstant}
G_{N}^{(10)}=8\pi^{6}g_{s}^{2} \ell_{s}^{8}\, .
\end{equation}

%%%%%%%%%%%%%%%%%%%%%%%%%%%%%%%%%%%%%%%%%%%%%%%%%%%%%%%%%%%%%%%%%%%%%%
%%%%%%%%%%%%%%%%%%%%%%%%%%%%%%%%%%%%%%%%%%%%%%%%%%%%%%%%%%%%%%%%%%%%%%
%%%%%%%%%%%%%%%%%%%%%%%%%%%%%%%%%%%%%%%%%%%%%%%%%%%%%%%%%%%%%%%%%%%%%%
%%%%%%%%%%%%%%%%%%%%%%%%%%%%%%%%%%%%%%%%%%%%%%%%%%%%%%%%%%%%%%%%%%%%%%
\section{Dimensional reduction on T$^{n}$ at zeroth order in $\alpha'$}
\label{sec-dimredO1}
%%%%%%%%%%%%%%%%%%%%%%%%%%%%%%%%%%%%%%%%%%%%%%%%%%%%%%%%%%%%%%%%%%%%%% 
%%%%%%%%%%%%%%%%%%%%%%%%%%%%%%%%%%%%%%%%%%%%%%%%%%%%%%%%%%%%%%%%%%%%%%
%%%%%%%%%%%%%%%%%%%%%%%%%%%%%%%%%%%%%%%%%%%%%%%%%%%%%%%%%%%%%%%%%%%%%%
%%%%%%%%%%%%%%%%%%%%%%%%%%%%%%%%%%%%%%%%%%%%%%%%%%%%%%%%%%%%%%%%%%%%%%

As a warm-up exercise (and also because of the recursive definition of the
action), we review the well-known dimensional reduction of the action at
zeroth order in $\alpha'$ using the Scherk-Schwarz formalism
\cite{Scherk:1979zr}. We add hats to all the 10-dimensional objects (fields,
indices, coordinates) and split the 10-dimensional world and tangent-space
indices as $(\hat{\mu})=(\mu,m)$ and $(\hat{a})=(a,i)$, with with
$\mu,\nu, \ldots$ and $a,b,\ldots =0,1,\cdots,9-n$ and $m,n,\ldots$ and
$i,j,\ldots = 1,\cdots,n$.

The Zehnbein and inverse-Zehnbein components $\hat{e}_{\hat{\mu}}{}^{\hat{a}}$
and $\hat{e}_{\hat{a}}{}^{\hat{\mu}}$ can be put in an upper-triangular form
by a local Lorentz transformation and, then, they can be decomposed in terms
of the $10-n$-dimensional Vielbein and inverse Vielbein components
$e_{\mu}{}^{a},e_{a}{}^{\mu}$, Kaluza-Klein (KK) vectors $A^{m}{}_{\mu}$ and
internal (T$^{n}$) metric Vielbeins and inverse Vielbein
$e_{m}{}^{i},e_{i}{}^{m}$

\begin{eqnarray}
\left( \hat{e}_{\hat{\mu}}{}^{\hat{a}} \right) =
\left(
\begin{array}{cc}
e_{\mu}{}^{a}
& 
A^{m}{}_{\mu} e_{m}{}^{i}
\\
& 
\\
0
& 
e_{m}{}^{i}
\\
\end{array}
\right)\!,
\hspace{.7cm}
\left( \hat{e}_{\hat{a}}{}^{\hat{\mu}} \right) =
\left(
\begin{array}{cc}
e_{a}{}^{\mu}
&
-A^{m}{}_{a}
\\
&
\\
0
&
e_{i}{}^{m}
\\
\end{array}
\right)\!.
\label{eq:KKbasisd}
\end{eqnarray}

\noindent
where $A^{m}{}_{a}= e_{a}{}^{\mu} A^{m}{}_{\mu}$. We will always assume that
all the $(10-n)$-dimensional fields with Lorentz indices are
$(10-n)$-dimensional world tensors contracted with the $(10-n)$-dimensional
Vielbeins. For instance, the field strengths of the KK vector fields
$F^{m}{}_{ab}$ are

\begin{equation}
\label{eq:KKfieldstrength}
F^{m}{}_{ab} = e_{a}{}^{\mu} e_{b}{}^{\nu} F^{m}{}_{\mu\nu}\, ,
\hspace{1cm}
F^{m}{}_{\mu\nu} \equiv 2\partial_{[\mu}A^{m}{}_{\nu]}\, .
\end{equation}

We denote the internal metric by

\begin{equation}
G_{mn} \equiv e_{m}{}^{i}e_{n}{}^{j} \delta_{ij}\, .
\end{equation}

The relation between the components of the  10-dimensional metric and
$(10-n)$-dimensional KK fields is

\begin{subequations}
  \label{eq:components10dmetric}
\begin{align}
  \hat{g}_{\mu\nu}
  & =
    g_{\mu\nu} -G_{mn}A^{m}{}_{\mu}A^{n}{}_{\nu}\, ,
  \\
  & \nonumber \\
  \hat{g}_{\mu m}
  & =
    -G_{mn}A^{n}{}_{\mu}\, ,
  \\
  & \nonumber \\
  \hat{g}_{mn}
  & =
    -G_{mn}\, .
\end{align}
\end{subequations}

The components of the 10-dimensional spin connection
$\hat{\omega}_{\hat{a}\hat{b}\hat{c}}$ decompose into those of the
$(10-n)$-dimensional one $\omega_{abc}$, the KK vector field strengths
$e_{im}F^{m}{}_{ab}$ and the pullback of the O$(n)$ connection 1-form $A^{ij}$
defined in Eq.~(\ref{eq:sonconnection}), as follows:

\begin{equation}
\begin{array}{rclrcl}
\hat{\omega}_{abc} & = & \omega_{abc}\, , & 
\hat{\omega}_{abi} & = & -\frac{1}{2}e_{im}F^{m}{}_{ab}\, , \\
& & & & & \\
\hat{\omega}_{ibc} & = & -\hat{\omega}_{bci}\, ,
\hspace{1cm}&
\hat{\omega}_{aij} & = & A^{ij}{}_{a}\, , \\
& & & & & \\
\hat{\omega}_{ibj} & = & -\frac{1}{2} e_{i}{}^{m} e_{j}{}^{n}
\partial_{b} G_{mn}\, , \hspace{1cm}\\
\end{array}
\end{equation}

\noindent
where we have used

\begin{equation}
  \label{eq:ede=eedG}
  e_{(i|}{}^{m}\partial_{a}e_{|m|j)}
  = -\tfrac{1}{2} e_{i}{}^{m}e_{j}{}^{n}\partial_{a} G_{mn}\, .
\end{equation}

Then, using the Palatini identity, it is not difficult to see that the first
two terms in the action Eq.~(\ref{heterotic}) take the following
$(10-n)$-dimensional form (up to a total derivative):

\begin{equation}
  \label{eq:EHdilreduction}
  \begin{aligned}
    \int d^{10}\hat{x}\sqrt{|\hat{g}|}\, e^{-2\hat{\phi}}\, \left\{ \hat{R}
      -4(\partial\hat{\phi})^{2} \right\}
        & \\
        & \\
        & \hspace{-5cm}
        =\int d^{n}z\int d^{10-n}x\sqrt{|g|}\,
        e^{-2\phi}\, \left\{ R -4(\partial\phi)^{2}
          -\tfrac{1}{4}\partial_{a}G_{mn}\partial^{a}G^{mn}
          -\tfrac{1}{4}G_{mn}F^{m}\cdot F^{n} \right\}\, ,
          \end{aligned}
\end{equation}

\noindent
where the $(10-n)$-dimensional dilaton field is related to the 10-dimensional
one by

\begin{equation}
\label{eq:components10ddilaton}
  \phi \equiv \hat{\phi} -\tfrac{1}{2}\log{|G|}\, ,
  \,\,\,\,\,\,\,
  |G|\equiv \text{det}(G_{mn})\, .
\end{equation}

At zeroth order in $\alpha'$, the last term that we have to reduce is the
kinetic term of the Kalb-Ramond 2-form $\sim (\hat{H}^{(0)})^{2}$. Following
Scherk and Schwarz, we consider the Lorentz components of the 3-form field
strength, because they are automatically combinations of gauge-invariant
objects. These are given in terms of the world-indices components by

\begin{subequations}
  \begin{align}
      \hat{H}_{ijk}
    &  =
      e_{i}{}^{m}e_{j}{}^{n}e_{k}{}^{p}\hat{H}_{mnp}\, ,
    \\
    & \nonumber \\
      \hat{H}_{aij}
& =
e_{i}{}^{m}e_{j}{}^{n}e_{a}{}^{\mu}
\left[ \hat{H}_{\mu mn} -A^{p}{}_{\mu}\hat{H}_{pmn}\right]\, ,
    \\
    & \nonumber \\
      \hat{H}_{abi}
& =
e_{i}{}^{m}e_{a}{}^{\mu}e_{b}{}^{\nu}
      \left[ \hat{H}_{\mu\nu m} -2A^{n}{}_{[\nu}\hat{H}_{\mu]nm}
      +A^{n}{}_{[\mu}A^{p}{}_{\nu]}\hat{H}_{npm }\right]\, ,
    \\
    & \nonumber \\
      \hat{H}_{abc}
& =
 e_{a}{}^{\mu}e_{b}{}^{\nu}e_{c}{}^{\rho}
\left[ \hat{H}_{\mu\nu\rho} -3A^{m}{}_{[\mu}\hat{H}_{\nu\rho]m}
  +3 A^{m}{}_{[\mu}A^{n}{}_{\nu}\hat{H}_{\rho]mn}
-A^{m}{}_{[\mu}A^{n}{}_{\nu}A^{p}{}_{\rho]}\hat{H}_{mnp}\right]\, ,
  \end{align}
\end{subequations}

\noindent
in general. At zeroth order in $\alpha'$, $\hat{H}^{(0)}{}_{mnp}=0$ and the
above expressions are simplified to 

\begin{subequations}
  \begin{align}
    \hat{H}^{(0)}{}_{ijk}
    & =
      0\, ,
    \\
    & \nonumber \\
    \hat{H}^{(0)}{}_{aij}
    & =
      e_{i}{}^{m}e_{j}{}^{n}e_{a}{}^{\mu}\hat{H}^{(0)}{}_{\mu mn}
      \nonumber \\
    & \nonumber \\
    & =
      e_{i}{}^{m}e_{j}{}^{n} \partial_{a}B^{(0)}{}_{mn}\, ,
    \\
    & \nonumber \\
    \hat{H}^{(0)}{}_{abi}
    & =
      e_{i}{}^{m}e_{a}{}^{\mu}e_{b}{}^{\nu}
      \left[ \hat{H}^{(0)}{}_{\mu\nu m}
      -2A^{n}{}_{[\nu}\hat{H}^{(0)}{}_{\mu]nm}\right]
      \nonumber \\
    & \nonumber \\
    & =
      e_{i}{}^{m}\left[G^{(0)}{}_{m\, ab} -B_{mn}F^{n}{}_{ab}\right]\, ,
    \\
    & \nonumber \\
    \hat{H}^{(0)}{}_{abc}
    & =
      e_{a}{}^{\mu}e_{b}{}^{\nu}e_{c}{}^{\rho}
      \left[ \hat{H}^{(0)}{}_{\mu\nu\rho} -3A^{m}{}_{[\mu}\hat{H}^{(0)}{}_{\nu\rho]m}
      +3 A^{m}{}_{[\mu}A^{n}{}_{\nu}\hat{H}^{(0)}{}_{\rho]mn}\right]
      \nonumber \\
    & \nonumber \\
    & =
      H^{(0)}{}_{abc}\, ,
  \end{align}
\end{subequations}

\noindent
where we have defined the potentials

\begin{subequations}
  \label{eq:KRreductionzerothorder}
  \begin{align}
    B_{mn}
    & \equiv
      \hat{B}^{(0)}{}_{mn}\, ,
      \\
    & \nonumber \\
    B^{(0)}{}_{m\, \mu}
    & \equiv
      \hat{B}_{\mu m} +\hat{B}_{mn}A^{n}{}_{\mu}\, ,
      \\
    & \nonumber \\
    B^{(0)}{}_{\mu\nu}
    & \equiv
      \hat{B}_{\mu \nu} +A^{m}{}_{[\mu} \hat{B}_{\nu]\, m}\, ,
  \end{align}
\end{subequations}

\noindent
and the field strengths

\begin{subequations}
  \begin{align}
    G^{(0)}{}_{m\, \mu\nu}
    & \equiv
      2\partial_{[\mu|}  B^{(0)}{}_{m\, |\nu]}\, ,
      \\
    & \nonumber \\
    H^{(0)}{}_{\mu\nu\rho}
    & \equiv
      3\partial_{[\mu}B^{(0)}{}_{\nu\rho]}
      -\tfrac{3}{2}A^{m}{}_{[\mu|}G^{(0)}{}_{m\, |\nu\rho]}
      -\tfrac{3}{2}B^{(0)}{}_{m\, [\mu} F^{m}{}_{\nu\rho]}\, .
  \end{align}
\end{subequations}

Then, the reduction of the Kalb-Ramond kinetic term gives

\begin{equation}
  \begin{aligned}
    \hat{H}^{(0)\, 2}
    & =
    H^{(0)\, 2} -3G^{mn}\left(G^{(0)}{}_{m}-B^{(0)}{}_{mp}F^{p}\right)\cdot
    \left(G^{(0)}{}_{n}-B^{(0)}{}_{nq}F^{q}\right)
    \\
    & \\
    & \hspace{.5cm}
    +3G^{mn}G^{pq}\partial_{a}B^{(0)}{}_{mp}\partial^{a}B^{(0)}{}_{nq}\, ,
  \end{aligned}
\end{equation}

\noindent
and, after integrating over the length of the compact coordinates $z^{m}$
($2\pi\ell_{s}$ by convention) it can be checked that the whole
$\mathcal{O}(1)$ action can be written in the compact form\footnote{The
  10-dimensional string coupling constant $g_{s}$ and Newton constant
  $G^{(10)}_{N}$ and the $(10-n)$-dimensional ones $g_{s}^{(10-n)}$ and $G^{(d)}_{N}$
  are related by
   \begin{subequations}
     \begin{align}
       g_{s}^{2} & = V_{n}/(2\pi\ell_{s})^{n}g_{s}^{(10-n)\, 2}\, , \\
                 & \nonumber \\
     \label{eq:relationsbetweenconstants}
        G_{N}^{(10)} & = G_{N}^{(10-n)} V_{n}\, ,
     \end{align}
   \end{subequations}
   where  $V_{n}$ is the volume of the $n$-dimensional compact
   space. Then,
   \begin{equation}
     \frac{g_{s}^{2}}{16\pi G_{N}^{(10)}}\int d^{n}z
     =
     \frac{g_{s}^{2}(2\pi\ell_{s})^{n}}{16\pi G_{N}^{(10)}}
    =
     \frac{g^{(10-n)\,2}_{s}}{16\pi G_{N}^{(10-n)}}\, .
   \end{equation}
}
\begin{equation}
  \label{eq:heterotic(10-n)order0}
  \begin{aligned}
    S^{(0)} & =  \frac{g^{(10-n)\,2}_{s}}{16\pi G_{N}^{(10-n)}}
    \int d^{10-n}x\sqrt{|g|}\, e^{-2\phi}\, \left\{ R -4(\partial\phi)^{2}
      -\tfrac{1}{8}\mathrm{Tr}\left(\partial_{a}M^{(0)\, -1}\partial^{a}M^{(0)}\right)
    \right.
    \\
    & \\
    &
    \left.
      -\tfrac{1}{4}\mathcal{F}^{(0)\, {\rm T}} M^{(0)\, -1}\cdot
      \mathcal{F}^{(0)}
      +\tfrac{1}{12}H^{(0)\, 2}\right\}\, ,
  \end{aligned}
\end{equation}

\noindent
where $M^{(0)}$ is the O$(n,n)$ matrix defined in Eq.~(\ref{eq:Mmatrixdef}) of
Appendix~\ref{app-Onn} with $B_{mn}$ replaced by $B^{(0)}{}_{mn}$ and where we
have defined the O$(n,n)$ vectors of 1-forms and 2-form field strengths

\begin{equation}
\mathcal{A}^{(0)}
  \equiv
  \left(
    \begin{array}{c}
      A^{m} \\ B^{(0)}{}_{m} 
    \end{array}
  \right)\, ,
  \hspace{1.5cm}
 \mathcal{F}^{(0)}
  \equiv
  \left(
    \begin{array}{c}
      F^{m} \\ G^{(0)}{}_{m} 
    \end{array}
    \right)\, .
\end{equation}
  
It is easy to show that $M^{(0)}$ is, indeed, a  O$(n,n)$ matrix

\begin{equation}
 M^{(0)}\Omega^{(0)} M^{(0)} \Omega^{(0)} = 1\, ,  
\end{equation}

\noindent
and rewrite the Kalb-Ramond field strength in the manifestly
O$(n,n)$-invariant form

\begin{equation}
  H^{(0)}{}_{\mu\nu\rho}
  =
  3\partial_{[\mu}B^{(0)}{}_{\nu\rho]}
  -\tfrac{3}{2}\mathcal{A}^{(0)\, {\rm T}}{}_{[\mu}\Omega^{(0)}\mathcal{F}^{(0)}{}_{\nu\rho]}\, .
\end{equation}

Actually, as it is well-known, the zeroth-order action $S^{(0)}$ given in
Eq.~(\ref{eq:heterotic(10-n)order0}) is manifestly invariant under O$(n,n)$
transformations which are understood as T-duality transformations from the
10-dimensional point of view.

%%%%%%%%%%%%%%%%%%%%%%%%%%%%%%%%%%%%%%%%%%%%%%%%%%%%%%%%%%%%%%%%%%%%%%
%%%%%%%%%%%%%%%%%%%%%%%%%%%%%%%%%%%%%%%%%%%%%%%%%%%%%%%%%%%%%%%%%%%%%%
%%%%%%%%%%%%%%%%%%%%%%%%%%%%%%%%%%%%%%%%%%%%%%%%%%%%%%%%%%%%%%%%%%%%%%
%%%%%%%%%%%%%%%%%%%%%%%%%%%%%%%%%%%%%%%%%%%%%%%%%%%%%%%%%%%%%%%%%%%%%%
\section{Dimensional reduction on T$^{n}$ with Yang-Mills fields and Heterotic 
Supergravity}
\label{sec-reductionheteroticsupergravity}
%%%%%%%%%%%%%%%%%%%%%%%%%%%%%%%%%%%%%%%%%%%%%%%%%%%%%%%%%%%%%%%%%%%%%% 
%%%%%%%%%%%%%%%%%%%%%%%%%%%%%%%%%%%%%%%%%%%%%%%%%%%%%%%%%%%%%%%%%%%%%%
%%%%%%%%%%%%%%%%%%%%%%%%%%%%%%%%%%%%%%%%%%%%%%%%%%%%%%%%%%%%%%%%%%%%%%
%%%%%%%%%%%%%%%%%%%%%%%%%%%%%%%%%%%%%%%%%%%%%%%%%%%%%%%%%%%%%%%%%%%%%%

In this section we are only going to take into account the addition of the
Yang-Mills fields which occur at first order in $\alpha'$, ignoring for the
moment the terms that involve the torsionful spin connection. This truncation,
which constitutes an intermediate step towards our final goal, is interesting
by itself because it corresponds to the bosonic sector of a theory with exact
local supersymmetry:\footnote{That is, the action is exactly invariant, not
  just up to terms of higher order in $\alpha'$.} $\mathcal{N}=1,d=10$
supergravity coupled to non-Abelian vector supermultiplets, also known as
\textit{Heterotic Supergravity}, constructed in
Ref.~\cite{Chapline:1982ww}. The action of this theory is

\begin{equation}
\label{eq:heteroticsupergravityaction}
S^{(h)}
=
\frac{g_{s}^{2}}{16\pi G_{N}^{(10)}}
\int d^{10}x\sqrt{|g|}\, 
e^{-2\phi}\, 
\left\{
R
-4(\partial\phi)^{2}
+\tfrac{1}{12}H^{(h)\, 2}
-\dfrac{\alpha'}{8}
F_{A}\cdot F^{A}
\right\}\, ,
\end{equation}

\noindent
where

\begin{equation}
  H^{(h)}{}_{\mu\nu\rho}
  =
  3\partial_{[\mu}B_{\nu\rho]}
+\frac{\alpha'}{4}\omega^{\rm YM}{}_{\mu\nu\rho}\, ,
\end{equation}

\noindent
and $F^{A}$ and $\omega^{\rm YM}$ are defined in Eqs.~(\ref{eq:FAdef}) and
(\ref{eq:oYMdef}), respectively.

Notice that the $\mathcal{O}(\alpha^{\prime\, 2})$ terms of this action have
to be kept in order to have exact local supersymmetry. 

The toroidal dimensional reduction of this theory in the case in which the
gauge group is Abelian was carried out along the same lines we are going to
follow here in Ref.~\cite{Chamseddine:1980cp,Maharana:1992my}. In the second
of these references the O$(n,n+n_{V})$ global symmetry of the resulting action
was related to the T-duality transformations of the Heterotic Superstring. In
the non-Abelian case, the gauge fields coming form the 10-dimensional gauge
fields cannot be rotated into Kaluza-Klein and winding vector fields coming
from the 10-dimensional metric and Kalb-Ramond fields. As a result,
O$(n,n+n_{V})$ is broken to  O$(n,n)$, or  O$(n,n+n_{A})$ where $n_{A}$ is the
number of Abelian gauge fields.

The reduction of the Einstein-Hilbert term and of the scalar kinetic term are
not modified by the inclusion of $\alpha'$ corrections. The definitions of
$(10-n)$-dimensional metric $g_{\mu\nu}$, dilaton $\phi$, KK vectors
$A^{m}{}_{\mu}$ and scalars $G_{mn}$ in terms of the 10-dimensional fields are
not modified by them either and they are still given by
Eqs.~(\ref{eq:components10dmetric}) and (\ref{eq:components10ddilaton}).

Because of the additional Yang-Mills Chern-Simons term in the Kalb-Ramond
field strength, we do expect modifications in the definitions of the
definitions of the $(10-n)$-fields that originate in the Kalb-Ramond 2-form,
namely the $(10-n)$-dimensional Kalb-Ramond 2-form $B^{(h)}_{\mu\nu}$, the
winding vectors $B^{(h)}{}_{m\, \mu}$, with respect to their zeroth-order
counterparts defined in Eqs.~(\ref{eq:KRreductionzerothorder}).\footnote{We
  use the superscript $(h)$ to indicate that these are the fields that arise
  in the reduction of Heterotic Supergravity and that possible contributions
  from the torsionful spin connection have not been taken into account.}

%%%%%%%%%%%%%%%%%%%%%%%%%%%%%%%%%%%%%%%%%%%%%%%%%%%%%%%%%%%%%%%%%%%%%%
%%%%%%%%%%%%%%%%%%%%%%%%%%%%%%%%%%%%%%%%%%%%%%%%%%%%%%%%%%%%%%%%%%%%%%
%%%%%%%%%%%%%%%%%%%%%%%%%%%%%%%%%%%%%%%%%%%%%%%%%%%%%%%%%%%%%%%%%%%%%%
%%%%%%%%%%%%%%%%%%%%%%%%%%%%%%%%%%%%%%%%%%%%%%%%%%%%%%%%%%%%%%%%%%%%%%
\subsection{Reduction of the Yang-Mills fields}
\label{sec-YMfields}
%%%%%%%%%%%%%%%%%%%%%%%%%%%%%%%%%%%%%%%%%%%%%%%%%%%%%%%%%%%%%%%%%%%%%% 
%%%%%%%%%%%%%%%%%%%%%%%%%%%%%%%%%%%%%%%%%%%%%%%%%%%%%%%%%%%%%%%%%%%%%%
%%%%%%%%%%%%%%%%%%%%%%%%%%%%%%%%%%%%%%%%%%%%%%%%%%%%%%%%%%%%%%%%%%%%%%
%%%%%%%%%%%%%%%%%%%%%%%%%%%%%%%%%%%%%%%%%%%%%%%%%%%%%%%%%%%%%%%%%%%%%%

It is convenient to start by studying the dimensional reduction of the
Yang-Mills fields. The Lorentz-indices decomposition of the gauge field is

\begin{subequations}
  \begin{align}
    \hat{A}^{A}{}_{i}
    & =
      e_{i}{}^{m}\hat{A}^{A}{}_{m}
      \equiv \varphi^{A}{}_{i}\, ,
      \\
    & \nonumber \\
    \hat{A}^{A}{}_{a}
    & =
      \hat{e}_{a}{}^{\mu} \hat{A}^{A}{}_{\mu}
      =
      e_{a}{}^{\mu}\left(\hat{A}^{A}{}_{\mu}-\hat{A}^{A}{}_{m}A^{m}{}_{\mu} \right)
      \equiv
      e_{a}{}^{\mu}A^{A}{}_{\mu}\, ,
  \end{align}
\end{subequations}

\noindent
which leads to the definition of the $(10-n)$-dimensional adjoint scalars
$\varphi^{A}{}_{i}$ and gauge vectors

\begin{subequations}
  \begin{align}
    \varphi^{A}{}_{i}
    & \equiv
      e_{i}{}^{m}\hat{A}^{A}{}_{m}\, ,
    \\
    & \nonumber \\
    A^{A}{}_{\mu}
    & \equiv
      \hat{A}^{A}{}_{\mu}-\hat{A}^{A}{}_{m}A^{m}{}_{\mu}\, .
  \end{align}
\end{subequations}

The components of 10-dimensional gauge field strength can be decomposed in
terms of these fields as follows:

\begin{subequations}
  \label{eq:YMreduction}
  \begin{align}
    \hat{F}^{A}{}_{ij}
    & =
      f^{A}{}_{BC}\varphi^{B}{}_{i}\varphi^{C}{}_{j}\, ,
    \\
    & \nonumber \\
    \hat{F}^{A}{}_{ai}
    & =
      \mathfrak{D}_{a}\varphi^{A}{}_{i}
      +\tfrac{1}{2}\varphi^{A}{}_{j}e^{j}{}_{m}e^{i}_{n}\partial_{a}G^{mn}\, ,
    \\
    & \nonumber \\
    \hat{F}^{A}{}_{ab}
    & =
      F^{A}{}_{ab} +\varphi^{A}{}_{i}e^{i}{}_{m}F^{m}{}_{ab}\, ,
  \end{align}
\end{subequations}

\noindent
where $F^{A}{}_{\mu\nu}$ is the standard Yang-Mills gauge field strength for
the $(10-n)$-dimensional gauge fields $A^{A}{}_{\mu}$ and $\mathfrak{D}$ is
the Yang-Mills and O$(n)$-covariant derivative

\begin{equation}
  \mathfrak{D}_{a}\varphi^{A}{}_{i}
  =
  \partial_{a}\varphi^{A}{}_{i}
  +f^{A}{}_{BC}A^{B}{}_{a}\varphi^{C}{}_{i}
  +  A^{ij}{}_{a}\varphi^{A}{}_{j}\, ,
\end{equation}

\noindent
where the SO$(n)$ composite connection is given in
Eq.~(\ref{eq:sonconnection}).

\begin{equation}
  \label{eq:FAreduction}
  \begin{aligned}
    \hat{F}_{A}\cdot \hat{F}^{A} =\,\, & F_{A}\cdot F^{A}
    +2\varphi^{A}{}_{i}e^{i}{}_{m}F^{m}\cdot F_{A} +
    \varphi^{A}{}_{i}\varphi_{A\, j}e^{i}{}_{m}e^{j}{}_{n}F^{m}\cdot F^{n}
    \\
    & \\
    & -\tfrac{1}{2}\varphi^{A}{}_{i}\varphi_{A\, j}e^{i}{}_{m}e^{j}{}_{n}G_{pq}
    \partial^{a}G^{mp}\partial_{a}G^{nq}
    -2\mathfrak{D}^{a}\varphi^{A}{}_{i}\mathfrak{D}_{a}\varphi_{A\, i}
    \\
    & \\
    &
    -2\mathfrak{D}^{a}\varphi^{A}{}_{i}\varphi_{A\, j}e^{i}{}_{m}e^{j}{}_{n}
    \partial_{a}G^{mn}
    +f_{ABC}f^{A}{}_{DE}\varphi^{B}{}_{i}\varphi^{D}{}_{i}
    \varphi^{C}{}_{j}\varphi^{E}{}_{j}\, .
  \end{aligned}
\end{equation}

Our next goal is the reduction of the Kalb-Ramond 3-form field strength. It is
convenient to start with the reduction of the Yang-Mills Chern-Simons term

\begin{subequations}
  \begin{align}
    \hat{\omega}^{YM}{}_{ijk}
    & =
      2f_{ABC} \varphi^{A}{}_{i}\varphi^{B}{}_{j}\varphi^{C}{}_{k}\, ,
    \\
    & \nonumber \\
    \hat{\omega}^{YM}{}_{aij}
    & =
      2\mathfrak{D}_{a}\varphi^{A}{}_{[i}\varphi_{A\, |j]}\, ,
    \\
    & \nonumber \\
    \hat{\omega}^{YM}{}_{abi}
    & =
      \left(2
      F^{A}{}_{ab}+\varphi^{A}{}_{j}e^{j}{}_{m}F^{m}{}_{ab}\right)
      \varphi_{A\, i} 
      -2e_{[a}{}^{\mu}e_{b]}{}^{\nu}e_{i}{}^{m}
      \partial_{\mu}\left(\hat{A}^{A}{}_{\nu}\hat{A}_{A\, m}\right)\, ,
    \\
    & \nonumber \\
    \hat{\omega}^{YM}{}_{abc}
    & =
      \omega^{YM}{}_{abc}
      +\tfrac{3}{2}\varphi_{A\, i}e^{i}{}_{m}A^{A}{}_{[a}F^{m}{}_{bc]}\, .
  \end{align}
\end{subequations}

%%%%%%%%%%%%%%%%%%%%%%%%%%%%%%%%%%%%%%%%%%%%%%%%%%%%%%%%%%%%%%%%%%%%%%
%%%%%%%%%%%%%%%%%%%%%%%%%%%%%%%%%%%%%%%%%%%%%%%%%%%%%%%%%%%%%%%%%%%%%%
%%%%%%%%%%%%%%%%%%%%%%%%%%%%%%%%%%%%%%%%%%%%%%%%%%%%%%%%%%%%%%%%%%%%%%
%%%%%%%%%%%%%%%%%%%%%%%%%%%%%%%%%%%%%%%%%%%%%%%%%%%%%%%%%%%%%%%%%%%%%%
\subsection{Reduction of the Kalb-Ramond field}
\label{sec-KRfield}
%%%%%%%%%%%%%%%%%%%%%%%%%%%%%%%%%%%%%%%%%%%%%%%%%%%%%%%%%%%%%%%%%%%%%% 
%%%%%%%%%%%%%%%%%%%%%%%%%%%%%%%%%%%%%%%%%%%%%%%%%%%%%%%%%%%%%%%%%%%%%%
%%%%%%%%%%%%%%%%%%%%%%%%%%%%%%%%%%%%%%%%%%%%%%%%%%%%%%%%%%%%%%%%%%%%%%
%%%%%%%%%%%%%%%%%%%%%%%%%%%%%%%%%%%%%%%%%%%%%%%%%%%%%%%%%%%%%%%%%%%%%%

Combining the results in the reduction of $\hat{H}^{(0)}$ with the reduction
of the Yang-Mills fields, we find

\begin{subequations}
  \label{eq:Hh1}
  \begin{align}
      \hat{H}^{(h)}{}_{ijk}
    & =
      % e_{i}{}^{m}e_{j}{}^{n}e_{k}{}^{p}\hat{H}^{(h)}{}_{mnp}
      % =
      % \frac{\alpha'}{4}\hat{\omega}^{YM}{}_{ijk}
      % =
      \frac{\alpha'}{2}f_{ABC} \varphi^{A}{}_{i}\varphi^{B}{}_{j}\varphi^{C}{}_{k}\, ,
    \\
    & \nonumber \\
    \hat{H}^{(h)}{}_{aij}
    & =
      e_{i}{}^{m}e_{j}{}^{n}\partial_{a}B_{mn}
      +\frac{\alpha'}{2}\mathfrak{D}_{a}\varphi^{A}{}_{[i|}\varphi_{A\, |j]}\, ,
    \\
    & \nonumber \\
    \hat{H}^{(h)}{}_{abi}
    & =
      e_{i}{}^{m}e_{[a}{}^{\mu}e_{b]}{}^{\nu}
      \left\{
      2\partial_{\mu}\left[\hat{B}_{\nu\, m} +B_{mn}A^{n}{}_{\nu}
      -\frac{\alpha'}{4}A^{A}{}_{\nu}\hat{A}_{A\, m} \right]
      \right.
    \nonumber \\
    & \nonumber \\
    &
      \hspace{.5cm}
     \left.
    -\left[B_{mn}
    -\frac{\alpha'}{4}\hat{A}^{A}{}_{m}\hat{A}_{A\, n}\right]F^{n}{}_{\mu\nu}
    +\frac{\alpha'}{2}\hat{A}_{A\, m}F^{A}{}_{\mu\nu}
    \right\}
    \\
    & \nonumber \\
    \hat{H}^{(h)}{}_{abc}
    & =
      e_{[a}{}^{\mu}e_{b}{}^{\nu}e_{c]}{}^{\rho}
      \left\{
      3\partial_{\mu}\left[\hat{B}_{\nu\rho}+A^{m}{}_{\nu}\hat{B}_{\rho m}
      +\frac{\alpha'}{4}\hat{A}_{A\, m}A^{m}{}_{\nu}A^{A}{}_{\rho} \right]
      \right.
    \nonumber \\
    & \nonumber \\
    &
      \hspace{.5cm}
      -3A^{m}{}_{\mu}\partial_{\nu}\left[\hat{B}_{\rho\, m} +B_{mn}A^{n}{}_{\rho}
      -\frac{\alpha'}{4}A^{A}{}_{\rho}\hat{A}_{A\, m} \right]
    \nonumber \\
    & \nonumber \\
    &
      \hspace{.5cm}
      \left.
     -3 \left[\hat{B}_{\mu\, m} +B_{mn}A^{n}{}_{\mu}
      -\frac{\alpha'}{4}A^{A}{}_{\mu}\hat{A}_{A\, m}
      \right]\partial_{\nu}A^{m}{}_{\rho}
      +\frac{\alpha'}{4}\omega^{YM}{}_{\mu\nu\rho}
      \right\}\, .
  \end{align}
\end{subequations}

This result suggests the following definitions of $(10-n)$-dimensional fields:

\begin{subequations}
  \begin{align}
    B^{(h)}{}_{mn}
    & \equiv
      \hat{B}_{mn}-\frac{\alpha'}{4}\hat{A}^{A}{}_{m}\hat{A}_{A\,n}\, ,
    \\
    & \nonumber \\
      B^{(h)}{}_{m\, \mu}
    & \equiv
\hat{B}_{\mu\, m} +B_{mn}A^{n}{}_{\mu}
      -\frac{\alpha'}{4}A^{A}{}_{\mu}\hat{A}_{A\, m}
      \nonumber \\
    & \nonumber \\
    & =
      \hat{B}_{\mu\, m}
      +\left(\hat{B}_{mn}-\frac{\alpha'}{4}\hat{A}^{A}{}_{m}\hat{A}_{A\,n}\right)
      \hat{g}^{np}\hat{g}_{p\mu}
      -\frac{\alpha'}{4}\hat{A}_{A\, m}\hat{A}^{A}{}_{\mu}\, ,
    \\
    & \nonumber \\
    B^{(h)}{}_{\mu\nu}
    & \equiv
    \hat{B}_{\mu\nu}+A^{m}{}_{[\mu}\hat{B}_{\nu] m}
      +\frac{\alpha'}{4}\hat{A}_{A\, m}A^{m}{}_{[\mu}A^{A}{}_{\nu]}
            \nonumber \\
    & \nonumber \\
    & =
      \hat{B}_{\mu\nu}+\hat{g}^{mn}\hat{g}_{m[\mu}\hat{B}_{\nu] n}
      +\frac{\alpha'}{4}\hat{A}_{A\, m}\hat{g}^{mn}\hat{g}_{n[\mu}\hat{A}^{A}{}_{\nu]}\, ,
  \end{align}
\end{subequations}

\noindent
and  $(10-n)$-dimensional field strengths

\begin{subequations}
\begin{align}
    G^{(h)}{}_{m\, \mu\nu}
    & \equiv
      2\partial_{[\mu|}B^{(h)}_{m\, |\nu]}\, ,
      \\
    & \nonumber \\
    H^{(h)}{}_{\mu\nu\rho}
    &  \equiv
      3\partial_{[\mu}B^{(h)}{}_{\nu\rho]}
      -\tfrac{3}{2}A^{m}{}_{[\mu|}G^{(h)}{}_{m\, |\nu\rho]}
      -\tfrac{3}{2}B^{(h)}{}_{m\,[\mu}F^{m}{}_{\nu\rho]}
      +\frac{\alpha'}{4}\omega^{YM}{}_{\mu\nu\rho}\, ,
\end{align}
\end{subequations}

\noindent
This allows us to rewrite the components of the Kalb-Ramond field strength in
the form

\begin{subequations}
  \label{eq:Hh2}
  \begin{align}
      \hat{H}^{(h)}{}_{ijk}
    & =
      % e_{i}{}^{m}e_{j}{}^{n}e_{k}{}^{p}\hat{H}^{(h)}{}_{mnp}
      % =
      % \frac{\alpha'}{4}\hat{\omega}^{YM}{}_{ijk}
      % =
      \frac{\alpha'}{2}f_{ABC} \varphi^{A}{}_{i}\varphi^{B}{}_{j}\varphi^{C}{}_{k}\, ,
    \\
    & \nonumber \\
    \hat{H}^{(h)}{}_{aij}
    & =
      e_{i}{}^{m}e_{j}{}^{n}\partial_{a}B^{(h)}{}_{mn}
      +\frac{\alpha'}{2}\mathfrak{D}_{a}\varphi^{A}{}_{i}\varphi_{A\, j}\, ,
    \\
    & \nonumber \\
    \hat{H}^{(h)}{}_{abi}
    & =
      e_{i}{}^{m} \left(G^{(h)}{}_{m\, ab} -B^{(h)}{}_{mn}F^{n}{}_{ab}\right)
    +\frac{\alpha'}{2}F^{A}{}_{ab}\varphi_{A\, i}\, ,
    \\
    & \nonumber \\
    \hat{H}^{(h)}{}_{abc}
    & =
    H^{(h)}{}_{abc}\, ,
  \end{align}
\end{subequations}

\noindent
so that the reduction of the kinetic term is 

\begin{equation}
  \label{eq:Hhreduction}
  \begin{aligned}
    \hat{H}^{(h)\, 2}
    & =
    H^{(h)\, 2}-3G^{mn}\left(G^{(h)}{}_{m} -B^{(h)}{}_{mp}F^{p}\right)\cdot
    \left(G^{(h)}{}_{n} -B^{(h)}{}_{nq}F^{q}\right)
    \\
    & \\
    & \hspace{.5cm}
    -\frac{3\alpha^{\prime\, 2}}{4}
    \varphi^{A}{}_{i}\varphi^{B}{}_{i} F_{A}\cdot F_{B} -3\alpha'
    e_{i}{}^{m}\varphi^{A}{}_{i} \left(G^{(h)}{}_{m}
      -B^{(h)}{}_{mn}F^{n}\right)\cdot F^{A}
    \\
    & \\
    & \hspace{.5cm}
    +3G^{mn}G^{pq}\partial^{a}B^{(h)}{}_{mp}\partial_{a}B^{(h)}{}_{nq}
    +\frac{3\alpha^{\prime\, 2}}{4} \varphi^{A}{}_{j} \varphi^{B}{}_{j}
    \mathfrak{D}^{a}\varphi_{A\, i}\mathfrak{D}_{a}\varphi_{B\, i}
    \\
    & \\
    & \hspace{.5cm}
    +3\alpha' \mathfrak{D}^{a}\varphi_{A\, i}\varphi^{A}{}_{j}
    e_{i}{}^{m}e_{j}{}^{n}\partial_{a}B^{(h)}{}_{mn}
-\frac{\alpha^{\prime\, 2}}{4}f_{ABC}f_{A'B'C'}
\varphi^{A}{}_{i}\varphi^{A'}{}_{i}
\varphi^{B}{}_{j}\varphi^{B'}{}_{j}
\varphi^{C}{}_{k}\varphi^{C'}{}_{k}\, .
  \end{aligned}
\end{equation}

Collecting all the terms (that is: Eqs.~(\ref{eq:EHdilreduction}),
(\ref{eq:FAreduction}) and (\ref{eq:Hhreduction})), we get

\begin{equation}
  \label{eq:EHdilreduction2}
  \begin{aligned}
    S^{(h)} & = \frac{g^{(10-n)\,2}_{s}}{16\pi G_{N}^{(10-n)}}
    \int d^{10-n}x\sqrt{|g|}\, e^{-2\phi}\, \left\{ R -4(\partial\phi)^{2} \right.
    \\
    & \\
    & -\tfrac{1}{4}\partial_{a}G_{mn}\partial^{a}G^{mn} +\frac{\alpha'}{16}
    \varphi^{A}{}_{i}\varphi_{A\, j}e^{i}{}_{m}e^{j}{}_{n}G_{pq}
    \partial^{a}G^{mp}\partial_{a}G^{nq}
    +\tfrac{1}{4}G^{mn}G^{pq}\partial^{a}B^{(h)}{}_{mp}\partial_{a}B^{(h)}{}_{nq}
    \\
    & \\
    & \hspace{.5cm}
    +\frac{\alpha'}{4}\mathfrak{D}^{a}\varphi^{A}{}_{i}\mathfrak{D}_{a}\varphi_{A\,
      i} +\frac{\alpha^{\prime\, 2}}{16} \varphi^{A}{}_{j} \varphi^{B}{}_{j}
    \mathfrak{D}^{a}\varphi_{A\, i}\mathfrak{D}_{a}\varphi_{B\, i}
    \\
    & \\
    & \hspace{.5cm} +\frac{\alpha'}{4} \mathfrak{D}^{a}\varphi_{A\,
      i}\varphi^{A}{}_{j} e_{i}{}^{m}e_{j}{}^{n}\partial_{a}B^{(h)}{}_{mn}
    +\frac{\alpha'}{4}\mathfrak{D}^{a}\varphi^{A}{}_{i}\varphi_{A\,
      j}e^{i}{}_{m}e^{j}{}_{n} \partial_{a}G^{mn}
    \\
    & \\
    & \hspace{.5cm}
    -\tfrac{1}{4}\left(G_{mn}-B^{(h)}{}_{mp}G^{pq}B^{(h)}{}_{qn}
      +\frac{\alpha'}{2}\varphi^{A}{}_{i}\varphi_{A\,
        j}e^{i}{}_{m}e^{j}{}_{n}\right) F^{m} \cdot F^{n}
    \\
    & \\
    & \hspace{.5cm}
    -\tfrac{1}{4}G^{mn}G^{(h)}{}_{m}\cdot G^{(h)}{}_{n}
    +\tfrac{1}{2}G^{mp}B^{(h)}{}_{pn}G^{(h)}{}_{m}\cdot F^{n}
    \\
    & \\
    & \hspace{.5cm}
    -\frac{\alpha'}{8}
    \left(K_{AB}+\frac{\alpha'}{2}\varphi_{A\, i}\varphi_{B\, i}\right)
    F^{A}\cdot F^{B}
    -\frac{\alpha'}{4} \varphi_{A\, i}e^{i}{}_{m}G^{mn} G^{(h)}{}_{n}\cdot F^{A}
    \\
    & \\
    & \hspace{.5cm} \left.
          -\frac{\alpha'}{4} \varphi_{A\, i}e^{i}{}_{m}
    \left(\delta^{m}{}_{n}-G^{mp}B^{(h)}{}_{pn}\right)F^{n}\cdot F^{A}
+\tfrac{1}{12}H^{(h)}\cdot H^{(h)} -V(\varphi)
    \right\}\, ,
\end{aligned}
\end{equation}

\noindent
where $K_{AB}$ is the Killing metric of the gauge group and we have defined
the scalar potential

\begin{equation}
  V(\varphi)
  \equiv
  \frac{\alpha'}{8}f_{ABC}f^{A}{}_{DE}\varphi^{B}{}_{i}\varphi^{D}{}_{i}
    \varphi^{C}{}_{j}\varphi^{E}{}_{j}
+\frac{\alpha^{\prime\, 2}}{48}f_{ABC}f_{A'B'C'}
\varphi^{A}{}_{i}\varphi^{A'}{}_{i}
\varphi^{B}{}_{j}\varphi^{B'}{}_{j}
\varphi^{C}{}_{k}\varphi^{C'}{}_{k}\, .        
\end{equation}

\noindent
Defining the scalar matrices

\begin{equation}
  G\equiv (G_{mn})\, ,
  \hspace{.5cm}
  B^{(h)}\equiv (B^{(h)}{}_{mn})\, ,
  \hspace{.5cm}
  \varphi \equiv (\varphi^{A}{}_{i}e^{i}{}_{m})\, ,
  \hspace{.5cm}
  K \equiv \frac{\alpha'}{2} (K_{AB})\, ,
\end{equation}

\noindent
and the O$(n,n+n_{V})$ vector of 2-form field strengths

\begin{equation}
  \mathcal{F}^{(h)}{}_{\mu\nu}
  \equiv
  \left(
    \begin{array}{c}
      F^{m}{}_{\mu\nu} \\ G^{(h)}{}_{m\, \mu\nu} \\ F^{A}{}_{\mu\nu} \\ 
    \end{array}
    \right)\, ,
\end{equation}

\noindent
we can rewrite their kinetic terms in the form 

\begin{equation}
  -\tfrac{1}{4} \mathcal{F}^{(h)\, {\rm T}}M^{(h)}\cdot\mathcal{F}^{(h)}\, ,
\end{equation}

\noindent
where $M^{(h)}$ is the symmetric matrix

\begin{equation}
  M^{(h)\,-1}
  \equiv
  \left(
    \begin{array}{ccc}
      G+B^{(h)\, {\rm T}}G^{-1}B^{(h)}+\varphi^{{\rm T}}K\varphi\hspace{.3cm}
      & -B^{(h)\, {\rm T}}G^{-1}\hspace{.3cm}
      & \left(\mathbb{1}_{n\times n}-B^{(h)\, {\rm T}}G^{-1}\right)\varphi^{{\rm T}}K \\
      & & \\
      -G^{-1}B^{(h)} & G^{-1} & G^{-1}\varphi^{{\rm T}}K \\
            & & \\
      K\varphi\left(\mathbb{1}_{n\times n} - G^{-1} B^{(h)}\right) & K\varphi G^{-1}
      & K +K\varphi G^{-1}\varphi^{{\rm T}}K \\
    \end{array}
  \right)\, .
\end{equation}

This is an O$(n,n+n_{V})$ matrix (for $n_{V}$ gauge fields) because it
satisfies

\begin{equation}
  M^{(h)} \Omega^{(h)} M^{(h)} \Omega^{(h)} = 1\, ,
  \,\,\,\,\,
  \text{with}
  \,\,\,\,\,
  \Omega^{(h)}
  \equiv
  \left(
    \begin{array}{ccc}
      0 & \mathbb{1}_{n\times n} & 0 \\
      \mathbb{1}_{n\times n} & 0 & 0 \\
      0 & 0 & -K \\
    \end{array}
    \right)\, ,
\end{equation}

\noindent
in a basis in which the Killing metric is not simply the identity. 

The kinetic terms of the scalar fields can be written in the form

\begin{equation}
-\tfrac{1}{8}\mathfrak{D}^{a}M^{(h)}\mathfrak{D}_{a}M^{^{(h)}\,-1}\, ,  
\end{equation}

\noindent
where the covariant derivative only acts on the gauge group.\footnote{There
  are no free SO$(n)$ indices in $M^{(h)}$.} The total action takes
a form very similar to the zeroth-order one
Eq.~(\ref{eq:heterotic(10-n)order0}):

\begin{equation}
  \label{eq:heterotic(10-n)orderh}
  \begin{aligned}
    S^{(h)} & =
    \frac{g^{(10-n)\,2}_{s}}{16\pi G_{N}^{(10-n)}}
    \int d^{10-n}x\sqrt{|g|}\, e^{-2\phi}\, \left\{ R -4(\partial\phi)^{2}
      -\tfrac{1}{8}\mathrm{Tr}\left(\mathfrak{D}^{a}M^{(h)}\mathfrak{D}_{a}M^{(h)\,-1}\right)
    \right.
    \\
    & \\
    &
    \left.
      -\tfrac{1}{4}\mathcal{F}^{(h)\, {\rm T}}M^{(h)\, -1}\cdot  \mathcal{F}^{(h)}
      +\tfrac{1}{12}H^{(h)\, 2}- V^{(h)}(\varphi)\right\}\, .
  \end{aligned}
\end{equation}

\noindent
At first sight, this action is formally O$(n,n+n_{V})$-invariant except for
the scalar potential. However, we cannot transform Abelian into non-Abelian
fields and \textit{vice-versa} and the O$(n,n+n_{V})$ invariance is broken in
the Chern-Simons term of $H^{(h)}$, in the kinetic term of the vector fields
and also in the kinetic terms of the scalars and therefore, generically, the
invariance is broken to just O$(n,n)$. If the gauge group is Abelian, the
scalar potential disappears, the covariant derivatives of the scalars can be
rewritten entirely in terms of partial derivatives and $H^{(h)}$ takes the
manifestly O$(n,n+n_{V})$-invariant form

\begin{equation}
  H^{(h)}{}_{\mu\nu\rho}
  =
  3\partial_{[\mu}B^{(h)}{}_{\nu\rho]}
  -\tfrac{3}{2}\mathcal{A}^{(h)\, {\rm T}}{}_{[\mu}\Omega^{(h)}\mathcal{F}^{(h)}{}_{\nu\rho]}\, .
\end{equation}

%%%%%%%%%%%%%%%%%%%%%%%%%%%%%%%%%%%%%%%%%%%%%%%%%%%%%%%%%%%%%%%%%%%%%%
%%%%%%%%%%%%%%%%%%%%%%%%%%%%%%%%%%%%%%%%%%%%%%%%%%%%%%%%%%%%%%%%%%%%%%
%%%%%%%%%%%%%%%%%%%%%%%%%%%%%%%%%%%%%%%%%%%%%%%%%%%%%%%%%%%%%%%%%%%%%%
%%%%%%%%%%%%%%%%%%%%%%%%%%%%%%%%%%%%%%%%%%%%%%%%%%%%%%%%%%%%%%%%%%%%%%
\section{Complete dimensional reduction on T$^{n}$ to
  \texorpdfstring{$\mathcal{O}(\alpha')$}{O(α')}}
\label{sec-completereductiontofirstorder}
%%%%%%%%%%%%%%%%%%%%%%%%%%%%%%%%%%%%%%%%%%%%%%%%%%%%%%%%%%%%%%%%%%%%%% 
%%%%%%%%%%%%%%%%%%%%%%%%%%%%%%%%%%%%%%%%%%%%%%%%%%%%%%%%%%%%%%%%%%%%%%
%%%%%%%%%%%%%%%%%%%%%%%%%%%%%%%%%%%%%%%%%%%%%%%%%%%%%%%%%%%%%%%%%%%%%%
%%%%%%%%%%%%%%%%%%%%%%%%%%%%%%%%%%%%%%%%%%%%%%%%%%%%%%%%%%%%%%%%%%%%%%

%%%%%%%%%%%%%%%%%%%%%%%%%%%%%%%%%%%%%%%%%%%%%%%%%%%%%%%%%%%%%%%%%%%%%%
%%%%%%%%%%%%%%%%%%%%%%%%%%%%%%%%%%%%%%%%%%%%%%%%%%%%%%%%%%%%%%%%%%%%%%
%%%%%%%%%%%%%%%%%%%%%%%%%%%%%%%%%%%%%%%%%%%%%%%%%%%%%%%%%%%%%%%%%%%%%%
%%%%%%%%%%%%%%%%%%%%%%%%%%%%%%%%%%%%%%%%%%%%%%%%%%%%%%%%%%%%%%%%%%%%%%
\subsection{Torsionful spin connection}
\label{sec-torsionfulspinconnection}
%%%%%%%%%%%%%%%%%%%%%%%%%%%%%%%%%%%%%%%%%%%%%%%%%%%%%%%%%%%%%%%%%%%%%% 
%%%%%%%%%%%%%%%%%%%%%%%%%%%%%%%%%%%%%%%%%%%%%%%%%%%%%%%%%%%%%%%%%%%%%%
%%%%%%%%%%%%%%%%%%%%%%%%%%%%%%%%%%%%%%%%%%%%%%%%%%%%%%%%%%%%%%%%%%%%%%
%%%%%%%%%%%%%%%%%%%%%%%%%%%%%%%%%%%%%%%%%%%%%%%%%%%%%%%%%%%%%%%%%%%%%%

The reduction of the terms involving the torsionful spin connection
$\hat{\Omega}^{(0)}_{(-)\, \hat{\mu}}{}^{\hat{a}\hat{b}}$ can be carried out
in two steps: first we just treat it as just another Yang-Mills field but with
the particular gauge group SO$(1,9)$. Then, we decompose the gauge group
indices into SO$(1,9-n)\times$SO$(n)$ indices.  As a matter of fact, we can
just take the results of the previous section and assume that the gauge group
has been extended to include SO$(1,9)$. 

Let us carry out the first step.

%%%%%%%%%%%%%%%%%%%%%%%%%%%%%%%%%%%%%%%%%%%%%%%%%%%%%%%%%%%%%%%%%%%%%%
%%%%%%%%%%%%%%%%%%%%%%%%%%%%%%%%%%%%%%%%%%%%%%%%%%%%%%%%%%%%%%%%%%%%%%
%%%%%%%%%%%%%%%%%%%%%%%%%%%%%%%%%%%%%%%%%%%%%%%%%%%%%%%%%%%%%%%%%%%%%%
%%%%%%%%%%%%%%%%%%%%%%%%%%%%%%%%%%%%%%%%%%%%%%%%%%%%%%%%%%%%%%%%%%%%%%
\subsubsection{First step}
\label{sec-firststep}
%%%%%%%%%%%%%%%%%%%%%%%%%%%%%%%%%%%%%%%%%%%%%%%%%%%%%%%%%%%%%%%%%%%%%% 
%%%%%%%%%%%%%%%%%%%%%%%%%%%%%%%%%%%%%%%%%%%%%%%%%%%%%%%%%%%%%%%%%%%%%%
%%%%%%%%%%%%%%%%%%%%%%%%%%%%%%%%%%%%%%%%%%%%%%%%%%%%%%%%%%%%%%%%%%%%%%
%%%%%%%%%%%%%%%%%%%%%%%%%%%%%%%%%%%%%%%%%%%%%%%%%%%%%%%%%%%%%%%%%%%%%%

As in the general Yang-Mills case, the reduction of the torsionful spin
connection gives two fields

\begin{subequations}
  \begin{align}
    \varphi^{\hat{a}\hat{b}}{}_{i}
    & \equiv
      e_{i}{}^{m}\hat{\Omega}^{(0)}_{(-)\, m}{}^{\hat{a}\hat{b}}\, ,
    \\
    & \nonumber \\
    A^{\hat{a}\hat{b}}{}_{\mu}
    & \equiv
      \hat{\Omega}^{(0)}_{(-)\, \mu}{}^{\hat{a}\hat{b}}
      -\hat{\Omega}^{(0)}_{(-)\, m}{}^{\hat{a}\hat{b}}A^{m}{}_{\mu}\, ,
  \end{align}
\end{subequations}
 
\noindent
and the different components of its curvature give

\begin{subequations}
  \label{eq:curvaturereduction}
  \begin{align}
    \hat{R}^{(0)\, \hat{a}\hat{b}}_{(-)}{}_{ij}
    & =
     -2\varphi^{\hat{a}}{}_{\hat{c}\, [i}\varphi^{\hat{c}\hat{b}}{}_{j]} \, ,
    \\
    & \nonumber \\
        \hat{R}^{(0)\, \hat{a}\hat{b}}_{(-)}{}_{ci}
    & =
      \widetilde{\mathfrak{D}}_{c}\varphi^{\hat{a}\hat{b}}{}_{i}
      +\tfrac{1}{2}\varphi^{\hat{a}\hat{b}}{}_{j}e^{j}{}_{m}e^{i}{}_{n}\partial_{c}G^{mn}\, ,
    \\
    & \nonumber \\
     \hat{R}^{(0)\, \hat{a}\hat{b}}_{(-)}{}_{cd}
    & =
      F^{\hat{a}\hat{b}}{}_{cd} +\varphi^{\hat{a}\hat{b}}{}_{i}e^{i}{}_{m}F^{m}{}_{cd}\, .
  \end{align}
\end{subequations}

\noindent
It is worth stressing that $F^{\hat{a}\hat{b}}{}_{cd}$ is the
$(10-n)$-dimensional curvature of the $(10-n)$-dimensional SO$(1,9)$ gauge
field $A^{\hat{a}\hat{b}}{}_{\mu}$

\begin{equation}
  \label{eq:Fabmn}
  F^{\hat{a}\hat{b}}{}_{\mu\nu}
  \equiv
  2\partial_{[\mu}A^{\hat{a}\hat{b}}{}_{\nu]}
  -2A^{\hat{a}}{}_{\hat{c}\,[\mu}A^{\hat{c}\hat{b}}{}_{\nu]}\, ,
\end{equation}

\noindent
and $\widetilde{\mathfrak{D}}_{c}$ is a
SO$(1,9)\times$SO$(n)$ covariant derivative with the same connection plus the
composite SO$(n)$ connection in Eq.~(\ref{eq:sonconnection}):

\begin{equation}
  \widetilde{\mathfrak{D}}_{c}\varphi^{\hat{a}\hat{b}}{}_{i}
  =
  \partial_{c}\varphi^{\hat{a}\hat{b}}{}_{i}
  -2A^{[\hat{a}|}{}_{\hat{d}}{}_{c}\varphi^{\hat{d}|\hat{b}]}{}_{i}
  +A^{ij}{}_{c}\varphi^{\hat{a}\hat{b}}{}_{j}\, .
\end{equation}

At this stage we can use the results of the previous section (the Heterotic
Supergravity case) to write the $(10-n)$-dimensional action that one gets
after completing the first step, because it has exactly the same form the same
form as that of the Heterotic Supergravity case
Eq.~(\ref{eq:heterotic(10-n)orderh}) if we define a new gauge index $X$ that
includes the 10-dimensional adjoint gauge group index $A$ and the adjoint
10-dimensional Lorentz index $[\hat{a}\hat{b}]$: $X=A,[\hat{a}\hat{b}]$. We
can write, directly and formally

\begin{equation}
  \label{eq:heterotic(10-n)1storder-1}
  \begin{aligned}
    S^{(1)} & =
    \frac{g^{(10-n)\,2}_{s}}{16\pi G_{N}^{(10-n)}}
    \int d^{10-n}x\sqrt{|g|}\, e^{-2\phi}\, \left\{ R -4(\partial\phi)^{2}
      -\tfrac{1}{8}\mathrm{Tr}\left(\widetilde{\mathfrak{D}}^{a}\widetilde{M}^{(1)}
      \widetilde{\mathfrak{D}}_{a}\widetilde{M}^{(1)\,-1}\right)
    \right.
    \\
    & \\
    & \hspace{.5cm}
    \left.
      -\tfrac{1}{4}\widetilde{\mathcal{F}}^{(1)\, {\rm T}}
      \widetilde{M}^{(1)\, -1}\cdot  \widetilde{\mathcal{F}}^{(1)}
      +\tfrac{1}{12}\widetilde{H}^{(1)\, 2}- V^{(1)}(\varphi)\right\}\, ,
  \end{aligned}
\end{equation}

\noindent
where the covariant derivative $\widetilde{\mathfrak{D}}$ only acts on $X$
indices,\footnote{With the connection $A^{A}$ on YM indices and with the
  connection $A^{\hat{a}\hat{b}}$ on SO$(1,9)$ indices.}
$\widetilde{\mathcal{F}}^{(1)}$ is the vector of $(10-n)$-dimensional 2-form
field strengths

\begin{equation}
  \widetilde{\mathcal{F}}^{(1)}
  \equiv
  \left(
    \begin{array}{c}
      F^{m} \\ G^{(1)}{}_{m} \\ F^{X}  \\ 
    \end{array}
  \right)\, ,
\,\,\,\,\,
\text{with}
\,\,\,\,\,
\left(F^{X}\right)
\equiv
\left(
\begin{array}{c}
F^{A} \\ F^{\hat{a}\hat{b}}  \\ 
\end{array}
\right)\, ,
\end{equation}

\noindent
where $F^{m}$ has been defined in Eq.~(\ref{eq:KKfieldstrength}), $F^{A}$ is
the field strength of $(10-n)$-dimensional Yang-Mills field,
$F^{\hat{a}\hat{b}}$ is the field strengths of the $(10-n)$-dimensional
SO$(1,9)$ gauge field defined in Eq.~(\ref{eq:Fabmn}) and

\begin{equation}
  G^{(1)}{}_{m\, \mu\nu}
  \equiv 2\partial_{[\mu|}B^{(1)}{}_{m\, |\nu]}\, .
\end{equation}

\noindent
$M^{(1)\,-1}$ is the matrix

\begin{equation}
  \label{eq:widetildeM1def}
  \widetilde{M}^{(1)\, -1}
  \equiv
  \left(
    \begin{array}{ccc}
      G+B^{(1)\, {\rm T}}G^{-1}B^{(1)}+\varphi^{{\rm T}}K\varphi\hspace{.3cm}
      & -B^{(1)\, {\rm T}}G^{-1}\hspace{.3cm}
      & \left(\mathbb{1}_{n\times n}-B^{(1)\, {\rm T}}G^{-1}\right)\varphi^{{\rm T}}K \\
      & & \\
      -G^{-1}B^{(1)} & G^{-1} & G^{-1}\varphi^{{\rm T}}K \\
            & & \\
      K\varphi\left(\mathbb{1}_{n\times n} - G^{-1} B^{(1)}\right) & K\varphi G^{-1}
      & K +K\varphi G^{-1}\varphi^{\rm T}K\\
    \end{array}
  \right)\, ,
\end{equation}

\noindent
where

\begin{equation}
  \begin{aligned}
    G & \equiv (G_{mn})\, , \hspace{.5cm} B^{(1)}\equiv (B^{(1)}{}_{mn})\, ,
    \hspace{.5cm} \varphi \equiv \left(\varphi^{X}{}_{i} e^{i}{}_{m}\right) =
    \left(\varphi^{A}{}_{i}e^{i}{}_{m},\varphi^{\hat{a}\hat{b}}{}_{i}e^{i}{}_{m}\right)\,
    ,
    \\
    & \\
    K &
    \equiv
    \frac{\alpha'}{2}\left(K_{XY}\right)
    =
    \frac{\alpha'}{2}
    \left(
      \begin{array}{cc}
        K_{AB} & 0 \\
        0 & -\hat{\eta}^{\hat{a}\hat{b}}{}_{\hat{c}\hat{d}}
      \end{array}
    \right)\, .
  \end{aligned}
\end{equation}

\noindent
Finally, the $(10-n)$-dimensional Kalb-Ramond field strength
$\widetilde{H}^{(1)}$ and the ``scalar potential''\footnote{The variables
  $\varphi^{A}{}_{i}$ become true $(10-n)$-dimensional scalars with O$(n)$ and
  adjoint gauge indices, but the variables $\varphi^{\hat{a}}{}_{\hat{b}\,i}$
  become both $(10-n)$-dimensional tensors and scalars, see
  Eqs.~(\ref{eq:varphiabred}).} $V^{(1)}$ are given by\footnote{We have
  neglected the $\mathcal{O}(\alpha^{\prime\,2})$ terms in $V^{(1)}$.}

\begin{eqnarray}
    \widetilde{H}^{(1)}{}_{\mu\nu\rho}
     &  \equiv & 
      3\partial_{[\mu}B^{(1)}{}_{\nu\rho]}
      -\tfrac{3}{2}A^{m}{}_{[\mu|}G^{(1)}{}_{m\, |\nu\rho]}
                 -\tfrac{3}{2}B^{(1)}{}_{m\,[\mu}F^{m}{}_{\nu\rho]}
                 \nonumber \\
     & & \nonumber \\
  & & 
      +\frac{\alpha'}{4}\left(\omega^{YM}{}_{\mu\nu\rho}
                 +\widetilde{\omega}^{L(0)}_{(-)\, \mu\nu\rho}\right)\, ,
  \\
     & \nonumber \\
V^{(1)}(\varphi)
  & \equiv & 
\frac{\alpha'}{8}\left(
    2\varphi^{\hat{a}}{}_{\hat{b}\,i}\varphi^{\hat{b}}{}_{\hat{c}\,j}
    \varphi^{\hat{c}}{}_{\hat{d}\, i}\varphi^{\hat{d}}{}_{\hat{a}\, j}
    -2\varphi^{\hat{a}}{}_{\hat{b}\,i}\varphi^{\hat{b}}{}_{\hat{c}\,i}
      varphi^{\hat{c}}{}_{\hat{d}\, j}\varphi^{\hat{d}}{}_{\hat{a}\, j}
             \right.
             \nonumber \\
     & & \nonumber \\
     & &
         \left.
         +f_{ABC}f^{A}{}_{DE}\varphi^{B}{}_{i}\varphi^{D}{}_{i}
      \varphi^{C}{}_{j}\varphi^{E}{}_{j}
    \right)\, .  
\end{eqnarray}

In the above expressions we have used the $(10-n)$-dimensional fields
$B^{(1)}{}_{mn},B^{(1)}{}_{m\, \mu}$ and $B^{(1)}{}_{\mu\nu}$. They are
defined in terms of the 10-dimensional ones by

\begin{subequations}
  \begin{align}
    B^{(1)}{}_{mn}
    & \equiv
      \hat{B}_{mn}-\frac{\alpha'}{4}
      \left(\hat{A}^{A}{}_{m}\hat{A}_{A\,n}
      +\hat{\Omega}^{(0)}_{(-)\,m}{}^{\hat{a}}{}_{\hat{b}}
      \hat{\Omega}^{(0)}_{(-)\,n}{}^{\hat{b}}{}_{\hat{a}}\right)\, ,
    \\
    & \nonumber \\
      B^{(1)}{}_{m\, \mu}
    & \equiv
      \hat{B}_{\mu\, m}
      +\left[\hat{B}_{mn}-\frac{\alpha'}{4}\left(\hat{A}^{A}{}_{m}\hat{A}_{A\,n}
      +\hat{\Omega}^{(0)}_{(-)\,m}{}^{\hat{a}}{}_{\hat{b}}
      \hat{\Omega}^{(0)}_{(-)\,n}{}^{\hat{b}}{}_{\hat{a}}\right)\right]
      \hat{g}^{np}\hat{g}_{p\mu}
      \nonumber \\
    & \nonumber \\
    & \hspace{.5cm}
      -\frac{\alpha'}{4}\left(\hat{A}_{A\, m}\hat{A}^{A}{}_{\mu}
      +\hat{\Omega}^{(0)}_{(-)\,m}{}^{\hat{a}}{}_{\hat{b}}
      \hat{\Omega}^{(0)}_{(-)\,\mu}{}^{\hat{b}}{}_{\hat{a}}
      \right)\, ,
    \\
    & \nonumber \\
    B^{(1)}{}_{\mu\nu}
    & \equiv
      \hat{B}_{\mu\nu}+\hat{g}^{mn}\hat{g}_{m[\mu}\hat{B}_{\nu] n}
-\frac{\alpha'}{4}\left(\hat{A}_{A\, m}\hat{A}^{A}{}_{[\mu|}
      +\hat{\Omega}^{(0)}_{(-)\,m}{}^{\hat{a}}{}_{\hat{b}}
      \hat{\Omega}^{(0)}_{(-)\,[\mu|}{}^{\hat{b}}{}_{\hat{a}}
      \right)
      \hat{g}^{mn}\hat{g}_{|\nu]n}\, .
  \end{align}
\end{subequations}

The action Eq.~(\ref{eq:heterotic(10-n)1storder-1}) contains implicitly
$\mathcal{O}(\alpha^{\prime\, 2})$ terms such as $H^{(1)\,2}$, as the original
action Eq.~(\ref{heterotic}), but it is convenient to keep them in order to
have more compact and gauge-invariant expressions. Eliminating all the
$\mathcal{O}(\alpha^{\prime\, 2})$ terms we would get an action which is gauge
invariant only to $\mathcal{O}(\alpha^{\prime})$, but this gauge invariance
and possible duality invariance (which we are going to discuss next) would not
be manifest.

Although the action Eq.~(\ref{eq:heterotic(10-n)1storder-1}) is apparently
manifestly O$(n,n)$-invariant, this is not so clear because the statement
assumes that all the terms not directly affected by the linear O$(n,n)$
transformations remain invariant. However, some of those terms, such as
$F^{\hat{a}}{}_{\hat{b}}\cdot F^{\hat{a}}{}_{\hat{b}}$, for instance, depend
on the internal Vielbein $e^{i}{}_{m}$ and/or the KK vectors $A^{m}{}_{\mu}$
which are not invariant. We have to move to the second phase and expand the
terms that depend on the fields with SO$(1,9)$ indices in terms of
$(10-n)$-dimensional fields.

%%%%%%%%%%%%%%%%%%%%%%%%%%%%%%%%%%%%%%%%%%%%%%%%%%%%%%%%%%%%%%%%%%%%%% 
%%%%%%%%%%%%%%%%%%%%%%%%%%%%%%%%%%%%%%%%%%%%%%%%%%%%%%%%%%%%%%%%%%%%%%
%%%%%%%%%%%%%%%%%%%%%%%%%%%%%%%%%%%%%%%%%%%%%%%%%%%%%%%%%%%%%%%%%%%%%%
%%%%%%%%%%%%%%%%%%%%%%%%%%%%%%%%%%%%%%%%%%%%%%%%%%%%%%%%%%%%%%%%%%%%%%
\subsubsection{Second step}
\label{sec-secondstep}
%%%%%%%%%%%%%%%%%%%%%%%%%%%%%%%%%%%%%%%%%%%%%%%%%%%%%%%%%%%%%%%%%%%%%% 
%%%%%%%%%%%%%%%%%%%%%%%%%%%%%%%%%%%%%%%%%%%%%%%%%%%%%%%%%%%%%%%%%%%%%%
%%%%%%%%%%%%%%%%%%%%%%%%%%%%%%%%%%%%%%%%%%%%%%%%%%%%%%%%%%%%%%%%%%%%%%
%%%%%%%%%%%%%%%%%%%%%%%%%%%%%%%%%%%%%%%%%%%%%%%%%%%%%%%%%%%%%%%%%%%%%%

The fields with SO$(1,9)$ indices $\varphi^{\hat{a}\hat{b}}{}_{i}$ and
$A^{\hat{a}\hat{b}}{}_{\mu}$ are further reduced as follows:

\begin{equation}
  \label{eq:varphiabred}
    \varphi^{\hat{a}\hat{b}}{}_{i}
     \longrightarrow
      \left\{
      \begin{array}{rcl}
        \varphi^{ab}{}_{i} & = &
                                 -\tfrac{1}{2}K^{(0)\, i\, ab}_{(+)}\, ,
        \\
        & & \\
        \varphi^{ai}{}_{j} & = &
                                 -P_{(-)}^{(0)\,a}{}_{ij}\, ,
       \\
                & & \\
        \varphi^{ij}{}_{k} & = & 0\, ,
        \\
      \end{array}
    \right.
\end{equation}

\noindent
and

\begin{equation}
  \label{eq:Aabred}
    A^{\hat{a}\hat{b}}{}_{\mu}
     \longrightarrow
      \left\{
      \begin{array}{rcl}
        A^{ab}{}_{\mu} & = & \Omega^{(0)}_{(-)\, \mu}{}^{ab}\, ,
        \\
                       & & \\
        A^{ai}{}_{\mu} & = & -\tfrac{1}{2}K^{(0)\,i}_{(-)\,\mu}{}^{a}\, ,
        \\
                       & & \\
        A^{ij}{}_{\mu} & = &
                             % A^{ij}{}_{\mu}-\tfrac{1}{2}e^{im}e^{jn}\partial_{\mu}B_{mn}
                             % \equiv
                             A^{(0)\,ij}_{(-)}{}_{\mu}\, ,
        \\
      \end{array}
    \right.
\end{equation}

\noindent
where the 2-form $K_{(-)}^{(0)\,i}{}_{\mu\nu}$, the O$(n)$ connection 1-form
$A^{(0)\, ij}_{(-)}$ and the Vielbein $P_{(-)}^{(0)}{}_{ij}$ we have been defined in
Eq.~(\ref{eq:Kpmdef}) with $B_{mn}$ replaced by $B^{(0)}{}_{mn}$.

Taking into account these expressions, the components of the
$(10-n)$-dimensional SO$(1,9)$ field strength $F^{\hat{a}\hat{b}}{}_{cd}$ that
occurs in the reduction of the curvature of the torsionful spin connection
Eq.~(\ref{eq:curvaturereduction}) are decomposed as follows:

\begin{subequations}
  \begin{align}
    F^{ab}{}_{\mu\nu}
    & =
      R^{(0)}_{(-)\, \mu\nu}{}^{ab}
      -\tfrac{1}{2}K_{(-)}^{(0)\,i}{}_{[\mu}{}^{a}K_{(-)}^{(0)\,i}{}_{\nu]}{}^{b}\, ,
    \\
    & \nonumber \\
    F^{ai}{}_{\mu\nu}
    & =
      -\mathfrak{D}_{(-)\, [\mu}K_{(-)}^{(0)\, i}{}_{\nu]}{}^{a}\, ,
    \\
    & \nonumber \\
    F^{ij}{}_{\mu\nu}
    & =
      F^{(0)\,ij}_{(-)}{}_{\mu\nu}
      +\tfrac{1}{2}K_{(-)}^{(0)\, i}{}_{[\mu}{}^{a}K_{(-)}^{(0)\, j}{}_{\nu]\,a}\, ,
  \end{align}
\end{subequations}

\noindent
where $\mathfrak{D}_{(-)}$ is the SO$(1,9-n)\times$O$(n)$ covariant
derivative with the ``$(0)(-)$'' connections, that is

\begin{equation}
  \mathfrak{D}_{(-)\, [\mu}K_{(-)}^{(0)\, i}{}_{\nu]}{}^{a}
  =
  \partial_{[\mu}K_{(-)}^{(0)\, i}{}_{\nu]}{}^{a}
  +A^{(0)\,ij}_{(-)}{}_{[\mu}K_{(-)}^{(0)\, j}{}_{\nu]}{}^{a}
   -\Omega^{(0)}_{(-)\, [\mu|}{}^{a}{}_{b}K_{(-)}^{(0)\, i}{}_{\nu]}{}^{b}\, .
 \end{equation}

 Now we use all these decompositions into
 $\widetilde{\omega}^{L(0)}_{(-)\, abc}$, the $(10-n)$-dimensional
 Lorentz Chern-Simons 3-form of the SO$(1,9)$ connection, obtaining

\begin{subequations}
   \begin{align}
   \widetilde{\omega}^{L(0)}_{(-)\, \mu\nu\rho}
   & =
   \omega^{L(0)}_{(-)\,  \mu\nu\rho}
   -\omega^{O(n)}_{(-)\,  \mu\nu\rho}
   +3  \mathfrak{D}_{(-)\,[\mu}K_{(-)}^{(0)\,i}{}_{\nu}{}^{a}K_{(-)}^{(0)\,i}{}_{\rho]\,a}\, ,
     \\
     & \nonumber \\
     \omega^{O(n)}_{(-)\,  \mu\nu\rho}
     & \equiv
       3F^{(0)\,ij}_{(-)}{}_{[\mu\nu|}A^{(0)\, ij}_{(-)}{}_{|\rho]}
       +2A^{(0)\,ij}_{(-)}{}_{[\mu|}A^{(0)\,jk}_{(-)}{}_{|\nu |}
       A^{(0)\,ki}_{(-)}{}_{|\rho]}\, .
   \end{align}
\end{subequations}

According to the discussion in Appendix~\ref{app-Onn} the 3-form
$\mathfrak{D}_{(-)\,[\mu}K^{(-)\,i}{}_{\nu}{}^{a}K^{(-)\,i}{}_{\rho]\,a}$ is
O$(n,n)$-invariant. Since it is also gauge and Lorentz-invariant it is
natural to eliminate it from the definition of the $(10-n)$-dimensional
Kalb-Ramond field strength $H^{(1)}$:

\begin{subequations}
  \begin{align}
    \widetilde{H}^{(1)}{}_{\mu\nu\rho}
    & =
      H^{(1)}{}_{\mu\nu\rho}
      +\frac{3\alpha'}{4}\mathfrak{D}_{(-)\,[\mu}K_{(-)}^{(0)\,i}{}_{\nu}{}^{a}
      K_{(-)}^{(0)\,i}{}_{\rho]\,a}\, ,
    \\
    & \nonumber \\
    H^{(1)}{}_{\mu\nu\rho}
     &  \equiv  
      3\partial_{[\mu}B^{(1)}{}_{\nu\rho]}
      -\tfrac{3}{2}A^{m}{}_{[\mu|}G^{(1)}{}_{m\, |\nu\rho]}
                 -\tfrac{3}{2}B^{(1)}{}_{m\,[\mu}F^{m}{}_{\nu\rho]}
        \nonumber  \\
     &  \nonumber \\
  & 
      +\frac{\alpha'}{4}\left(\omega^{YM}{}_{\mu\nu\rho}
                 +\omega^{L(0)}_{(-)\,  \mu\nu\rho} -\omega^{O(n)}_{(-)\,  \mu\nu\rho}
                 \right)\, .
  \end{align}
\end{subequations}

Observe that the Chern-Simons 3-form of the composite O$(n)_{(-)}$ connection
$A^{(0)\, ij}_{(-)}$ is not O$(n,n)$-invariant, because it is not invariant
under the compensating local O$(n)_{(-)}$ transformation (it is invariant up
to a total derivative, as any Chern-Simons 3-form).\footnote{The presence of
  this term has been observed in Ref.~\cite{Eloy:2019hnl}} It can be
compensated with a standard Nicolai-Townsend transformation of the
$(10-n)$-dimensional 2-form $B^{(1)}$,\footnote{The Nicolai-Townsend
  transformations were first found in Ref.~\cite{Nicolai:1980td} in a
  different context. They were shown to be necessary in the coupling of vector
  multiplets to $\mathcal{N}=1,d=10$ supergravity in
  \cite{Bergshoeff:1981um}.} though. Thus, as different from the $n=1$ case,
this field is now not T-duality invariant to first order in $\alpha'$. Observe
that these transformations do not affect $B^{(0)}$, which is the field that
occurs in $\Omega^{(0)}_{(-)\, \mu}{}^{a}{}_{b}$ and $A^{(0)\,ij}_{(-)}$.

Next, let us consider the decomposition of the "scalar potential"
$V^{(1)}(\varphi)$.  Using Eqs.~(\ref{eq:varphiabred}), we get

\begin{equation}
  \begin{aligned}
    V^{(1)}(\varphi) & =
    \frac{\alpha'}{8} \left\{
      \tfrac{1}{8}K^{(0)\,i}_{(+)\,ab}K^{(0)\,i}_{(+)\,cd}
      K^{(0)\,j\,ac}_{(+)}K^{(0)\,j\,bd}_{(+)}
      +\tfrac{1}{2}K^{(0)\,i}_{(+)\, ab}K^{(0)\,j}_{(+)\,cd}
      K^{(0)\,i\,ac}_{(+)}K^{(0)\,j\,bd}_{(+)}
    \right.
    \\
    & \\
    & \hspace{.5cm}
\left.
    % -2^{6}P_{(-)}^{(0)\,a}{}_{ki}P_{(-)}^{(0)}{}_{a\,li}P_{(-)}^{(0)\,b}{}_{ki}P_{(-)}^{(0)}{}_{b\,li}
    % +2^{6}P_{(-)}^{(0)\,a}{}_{ki}P_{(-)}^{(0)}{}_{a\,lj}P_{(-)}^{(0)\,b}{}_{li}P_{(-)}^{(0)}{}_{b\,kj}
+F_{(-)}^{(0)\,ij}\cdot F_{(-)}^{(0)\,ij}
      +f_{ABC}f^{A}{}_{DE}\varphi^{B}{}_{i}\varphi^{D}{}_{i}
      \varphi^{C}{}_{j}\varphi^{E}{}_{j} \right\}\, ,
  \end{aligned}
\end{equation}

\noindent
where we have used the identity Eq.~(\ref{eq:FFPPPP}). All terms are
manifestly O$(n,n)$-invariant.

Next, we have to decompose the kinetic term for the vector field strengths
into its O$(n,n)$ invariant form and the rest:

\begin{equation}
  \begin{aligned}
    \widetilde{\mathcal{F}}^{(1)\,{\rm T}} \widetilde{M}^{(1)\,-1}\cdot
    \widetilde{\mathcal{F}}^{(1)}
    & =
    \mathcal{F}^{(1)\, {\rm T}} M^{(1)\,-1}\cdot \mathcal{F}^{(1)}
    +\frac{\alpha'}{2}\left[F_{A}\cdot F^{A} +R^{(0)}_{(-)}{}^{a}{}_{b}\cdot
      R^{(0)}_{(-)}{}^{b}{}_{a}
      \right.
    \\
    & \\
    &\hspace{.5cm}
    +R^{(0)}_{(-)\,abcd}\left(K_{(-)}^{(0)\,i\,ac}K_{(-)}^{(0)\,i\,bd}
      +K_{(+)}^{(0)\,i\,ab}K_{(+)}^{(0)\,i\,cd}\right)
    \\
    & \\
    &\hspace{.5cm}
    +\tfrac{1}{4}K_{(-)\,ab}^{(0)\,i}K_{(-)\,cd}^{(0)\,i}
    K_{(-)}^{(0)\,j\,ac}K_{(-)}^{(0)\,j\,bd}
    -\tfrac{1}{8}\left(K_{(-)}^{(0)\,i}\cdot K_{(-)}^{(0)\,j}\right)
    \left(K_{(-)}^{(0)\,i}\cdot K_{(-)}^{(0)\,j}\right)
    \\
    & \\
    &\hspace{.5cm}
    -\tfrac{1}{8}
    K_{(-)\,ab}^{(0)\,i} K_{(-)\,cd}^{(0)\,j}K_{(-)}^{(0)\,i\,ac} K_{(-)}^{(0)\,j\,bd}
-\tfrac{1}{2}K_{(+)\,ab}^{(0)\,i}K_{(+)\,cd}^{(0)\,i}K_{(-)}^{(0)\,j\,ac}K_{(-)}^{(0)\,j\,bd}
    \\
    & \\
    &\hspace{.5cm} +2\mathfrak{D}_{(-)\, [\mu}K_{(-)}^{(0)\, i}{}_{\nu]}{}^{a}
    \mathfrak{D}^{(0)\,[\mu|}_{(-)}K_{(-)}^{(0)\, i\,|\nu]}{}_{a}
    -F^{(0)\,ij}_{(-)}\cdot F^{(0)\,ij}_{(-)}
    -F^{(0)\,ij\,\mu\nu}_{(-)}K_{(-)}^{(0)\,i}{}_{\mu}{}^{a}K_{(-)}^{(0)\,j}{}_{\nu\,a}
    \\
    & \\
    &\hspace{.5cm}
    \left.
    +2\varphi_{A\,i} F^{A}\cdot  K_{(+)}^{(0)\,i}
+4P_{(-)\,\,ji}^{(0)\,a}\mathfrak{D}^{(0)}_{(-)\, \mu}K^{(0)\,j}_{(-)\,\nu
  a}K_{(+)}^{(0)\,i\,\mu\nu}\right]\, ,
  \end{aligned}
\end{equation}

\noindent
\noindent
where the covariant derivative $\mathfrak{D}_{(-)}^{(0)}$ contains the
O$(n)_{(-)}$ connection $A_{(-)}^{(0)\,ij}$ and the torsionful Lorentz
connection $\Omega^{(0)}_{(-)\,\mu}{}^{ab}$, and where, now, the matrix
$M^{(1)\,-1}$ is given by the upper-left corner $2\times 2$ part of
$\widetilde{M}^{(1)\, -1}$ in Eq.~(\ref{eq:widetildeM1def}), namely

\begin{equation}
  M^{(1)\,-1}
  \equiv
  \left(
    \begin{array}{cc}
      G+B^{(1)\, {\rm T}}G^{-1}B^{(1)}+\varphi^{{\rm T}}K\varphi\hspace{.3cm}
      & -B^{(1)\, {\rm T}}G^{-1}\hspace{.3cm}\\
       & \\
      -G^{-1}B^{(1)} & G^{-1} \\
    \end{array}
  \right)\, ,
\end{equation}

\noindent
and

\begin{equation}
  \mathcal{F}^{(1)}
  \equiv
  \left(
    \begin{array}{c}
      F^{m} \\ G^{(1)}{}_{m}\\
    \end{array}
    \right)\, .
\end{equation}

Finally, let us consider the scalar's kinetic term.  Expanding this term,
reducing and keeping only terms of first order in $\alpha'$ we arrive at

\begin{equation}
  \begin{aligned}
    \mathrm{Tr}\left(\widetilde{\mathfrak{D}}^{a}\widetilde{M}^{(1)}
      \widetilde{\mathfrak{D}}_{a}\widetilde{M}^{(1)\,-1}\right)
    & =
\mathrm{Tr}\left(\partial^{a}M^{(1)}\partial_{a}M^{(1)\,-1}\right)
-\alpha' \mathfrak{D}_{(-)}^{(0)\,c}
\left(\mathcal{U}_{(+)}{}^{i}\varphi_{A\,i}\right)^{T}
\Omega^{(1)}  \mathfrak{D}^{(0)}_{(-)\,c}\left(\mathcal{U}_{(+)}{}^{j}\varphi^{A}{}_{j}\right)
    \\
    & \\
    &
        \hspace{.5cm}
-\frac{\alpha'}{4} \mathfrak{D}_{(-)}^{(0)\,c}
\left(\mathcal{U}_{(+)}{}^{i}K_{(+)\,i}^{(0)}{}^{a}{}_{b}\right)^{T}
\Omega^{(1)} \mathfrak{D}^{(0)}_{(-)\,c}
\left(\mathcal{U}_{(+)}{}^{j}K_{(+)\,j}^{(0)}{}^{b}{}_{a}\right)
    \\
    & \\
    &
        \hspace{.5cm}
-2\alpha' \mathfrak{D}_{(-)}^{(0)\,c}
\left(\mathcal{U}_{(+)}{}^{i}P_{(+)}^{(0)\,a}{}_{ij}\right)^{T}
\Omega^{(1)}  \mathfrak{D}^{(0)}_{(-)\,c}\left(\mathcal{U}_{(+)}{}^{k}P_{(+)\,a\,kj}^{(0)}\right)\, .
  \end{aligned}
\end{equation}

Finally, combining all the partial results we get the final,
manifestly-O$(n,n)$-invariant form of the $(10-n)$-dimensional action

\begin{equation}
  \label{eq:dimredaction1}
  \begin{aligned}
    S^{(1)} & = \frac{g^{(10-n)\,2}_{s}}{16\pi G_{N}^{(10-n)}}
    \int d^{10-n}x\sqrt{|g|}\, e^{-2\phi}\, \left\{ R -4(\partial\phi)^{2}
        -\tfrac{1}{8}\mathrm{Tr}\left(\partial^{a}M^{(1)}\partial_{a}M^{(1)\,-1}\right)
\right.
    \\
    & \\
    &
    -\tfrac{1}{4}\mathcal{F}^{(1)\, {\rm T}} M^{(1)\,-1}\cdot
    \mathcal{F}^{(1)}
        +\tfrac{1}{12}\widetilde{H}^{(1)\,2}
    -\frac{\alpha'}{8}\left[F_{A}\cdot F^{A}
      +R^{(0)}_{(-)}{}^{a}{}_{b}\cdot R^{(0)}_{(-)}{}^{b}{}_{a}
        \right.
    \\
    & \\
    &\hspace{.5cm}
    +R^{(0)}_{(-)\,abcd}\left(K_{(-)}^{(0)\,i\,ac}K_{(-)}^{(0)\,i\,bd}
      +K_{(+)}^{(0)\,i\,ab}K_{(+)}^{(0)\,i\,cd}\right)
    \\
    & \\
    &\hspace{.5cm}
    +\tfrac{1}{4}K_{(-)\,ab}^{(0)\,i}K_{(-)\,cd}^{(0)\,i}
    K_{(-)}^{(0)\,j\,ac}K_{(-)}^{(0)\,j\,bd}
    -\tfrac{1}{8}\left(K_{(-)}^{(0)\,i}\cdot K_{(-)}^{(0)\,j}\right)
    \left(K_{(-)}^{(0)\,i}\cdot K_{(-)}^{(0)\,j}\right)
    \\
    & \\
    &\hspace{.5cm}
    -\tfrac{1}{8}
    K_{(-)\,ab}^{(0)\,i} K_{(-)\,cd}^{(0)\,j}K_{(-)}^{(0)\,i\,ac} K_{(-)}^{(0)\,j\,bd}
-\tfrac{1}{2}K_{(+)\,ab}^{(0)\,i}K_{(+)\,cd}^{(0)\,i}K_{(-)}^{(0)\,j\,ac}K_{(-)}^{(0)\,j\,bd}
    \\
    & \\
    &\hspace{.5cm}
+\tfrac{1}{8} K^{(0)\,i}_{(+)\,ab}K^{(0)\,i}_{(+)\,cd}
      K^{(0)\,j\,ac}_{(+)}K^{(0)\,j\,bd}_{(+)}
      +\tfrac{1}{8}K^{(0)\,i}_{(+)\, ab}K^{(0)\,j}_{(+)\,cd}
      K^{(0)\,i\,ac}_{(+)}K^{(0)\,j\,bd}_{(+)}
    \\
    & \\
    &\hspace{.5cm} +2\mathfrak{D}^{(0)}_{(-)\, [\mu}K_{(-)}^{(0)\, i}{}_{\nu]}{}^{a}
    \mathfrak{D}^{(0)\,[\mu|}_{(-)}K_{(-)}^{(0)\, i\,|\nu]}{}_{a}
    % -F^{(0)\,ij}_{(-)}\cdot F^{(0)\,ij}_{(-)}
    -F^{(0)\,ij\,\mu\nu}_{(-)}K_{(-)}^{(0)\,i}{}_{\mu}{}^{a}K_{(-)}^{(0)\,j}{}_{\nu\,a}
    \\
    & \\
    &\hspace{.5cm}
    +2\varphi_{A\,i} F^{A}\cdot  K_{(+)}^{(0)\,i}
+4P_{(-)\,\,ji}^{(0)\,a}\mathfrak{D}^{(0)}_{(-)\, \mu}K^{(0)\,j}_{(-)\,\nu
  a}K_{(+)}^{(0)\,i\,\mu\nu}
    \\
    & \\
    &\hspace{.5cm}
-\mathfrak{D}^{(0)\,c}_{(-)}
\left(\mathcal{U}_{(+)}{}^{i}\varphi_{A\,i}\right)^{T}
\Omega^{(1)}  \mathfrak{D}^{(0)}_{(-)\,c}
\left(\mathcal{U}_{(+)}{}^{j}\varphi^{A}{}_{j}\right)
    \\
    & \\
    &\hspace{.5cm}
-2  \mathfrak{D}^{(0)\,c}_{(-)}
\left(\mathcal{U}_{(+)}{}^{i}P_{(+)}^{(0)\,a}{}_{ij}\right)^{T}
\Omega^{(1)}
 \mathfrak{D}^{(0)}_{(-)\,c}\left(\mathcal{U}_{(+)}{}^{k}P_{(+)\,a\,kj}^{(0)}\right)
    \\
    & \\
    &\hspace{.5cm}
    \left.
-\tfrac{1}{4}\mathfrak{D}_{(-)}^{(0)\,c}
\left(\mathcal{U}_{(+)}{}^{i}K_{(+)\,i}^{(0)}{}^{a}{}_{b}\right)^{T}
\Omega^{(1)} \mathfrak{D}^{(0)}_{(-)\,c}\left(\mathcal{U}_{(+)}{}^{j}K_{(+)\,j}^{(0)}{}^{b}{}_{a}\right)
\right]
    \\
    & \\
    & \hspace{.5cm}
\left.
    \left.
% +F_{(-)}^{(0)\,ij}\cdot F_{(-)}^{(0)\,ij}
      +f_{ABC}f^{A}{}_{DE}\varphi^{B}{}_{i}\varphi^{D}{}_{i}
      \varphi^{C}{}_{j}\varphi^{E}{}_{j} 
    \right]\right\}\, .
\end{aligned}
\end{equation}

This dimensionally-reduced action has two main immediate uses: the derivation
of equations of motion and the derivation of an entropy formula. The equations
of motion are very complicated and involve, in principle, derivatives higher
than 2. According to the lemma proven in Ref.~\cite{Bergshoeff:1989de}, the
terms of higher-order in derivatives (possibly in combination with some terms
of lower order) are, actually, proportional to $\alpha'$ and combinations of
the zeroth-order equations of motion and can, in be disregarded in
practice. However, in order to find all the terms which can be ignored one has
to use the lower-dimensional version of the lemma, which requires the
identification of all the fields which originate, purely, on the
10-dimensional torsionful spin connection. Therefore, it is far easier to deal
with the 10-dimensional equations of motion and perform the dimensional
reduction of the solution using the rules derived in this paper.

The entropy formula, though, can be readily obtained and used in $(10-n)$
dimensions, as we are going to show in the next section.

%%%%%%%%%%%%%%%%%%%%%%%%%%%%%%%%%%%%%%%%%%%%%%%%%%%%%%%%%%%%%%%%%%%%%% 
%%%%%%%%%%%%%%%%%%%%%%%%%%%%%%%%%%%%%%%%%%%%%%%%%%%%%%%%%%%%%%%%%%%%%%
%%%%%%%%%%%%%%%%%%%%%%%%%%%%%%%%%%%%%%%%%%%%%%%%%%%%%%%%%%%%%%%%%%%%%%
%%%%%%%%%%%%%%%%%%%%%%%%%%%%%%%%%%%%%%%%%%%%%%%%%%%%%%%%%%%%%%%%%%%%%%
\section{Entropy formula}
\label{sec-entropyformula}
%%%%%%%%%%%%%%%%%%%%%%%%%%%%%%%%%%%%%%%%%%%%%%%%%%%%%%%%%%%%%%%%%%%%%% 
%%%%%%%%%%%%%%%%%%%%%%%%%%%%%%%%%%%%%%%%%%%%%%%%%%%%%%%%%%%%%%%%%%%%%%
%%%%%%%%%%%%%%%%%%%%%%%%%%%%%%%%%%%%%%%%%%%%%%%%%%%%%%%%%%%%%%%%%%%%%%
%%%%%%%%%%%%%%%%%%%%%%%%%%%%%%%%%%%%%%%%%%%%%%%%%%%%%%%%%%%%%%%%%%%%%%

If we change the index $10$ by $D$, the action Eq.~(\ref{heterotic}) is
identical to the action one would obtain by trivial dimensional reduction of
the 10-dimensional theory on a $m\equiv(10-D)$-dimensional torus. Then, with
the same change, Eq.~(\ref{eq:dimredaction1}) is the action obtained by a
fully non-trivial dimensional reduction on T$^{n}$, from $D$ to $d=D-n=10-n-m$
dimensions. The solutions of the equations of motion of this $d$-dimensional
theory are solutions which, uplifted to $d+n=D$ dimensions with the rules
derived in this paper, are, then, solutions of the equations of motion of the
original action Eq.~(\ref{heterotic}) with $10$ replaced by $D$ and which can
be trivially uplifted again to $10$ dimensions.

An important example of solutions of this type are the heterotic version of
the 4-dimensional, 4-charge extremal black holes whose microscopic entropy was
originally computed in Refs.~\cite{Maldacena:1996gb,Johnson:1996ga} (see also
Ref.~\cite{Kaplan:1996ev})\footnote{These 4-dimensional solutions can be
  embedded either in the Heterotic or in type~II theories. They were first
  found in Ref.~\cite{Cvetic:1995uj} and further discussed in
  Ref.~\cite{Cvetic:1995yq,Cvetic:1995bj}. They were rediscovered in
  Ref.~\cite{Rahmfeld:1995fm} in the context of the so-called STU model
  \cite{Duff:1995sm} in a form that made it easier to identify harmonic
  functions and charges to 10-dimensional string theory extended objects.}
and whose first-order $\alpha'$ corrections were recently computed in
Refs.~\cite{Chimento:2018kop,Cano:2018brq}.\footnote{In these references the
  first-order in $\alpha'$ corrections to the complete geometry were computed.
  Earlier work in which only the corrections to the near-horizon geometry were
  computed and then, used to compute the corrections to the entropy can be
  found in Refs.~\cite{Behrndt:1998eq, LopesCardoso:1998tkj,
    LopesCardoso:1999cv, LopesCardoso:1999fsj, Sahoo:2006pm,
    Prester:2008iu}. Some of the drawbacks of these methods, such as the
  problem of identification of asymptotic charges and the possible
  incompleteness of the higher-order terms considered have been discussed in
  Ref.~\cite{Cano:2018brq,Faedo:2019xii}.}

If these 4-dimensional solutions are uplifted to $6$ dimensions ($n=2$), they
are solutions of our original action Eq.~(\ref{heterotic}) with $D=6$
($m=4$). Thus, they are solutions of the equations of motion of the action
Eq.~(\ref{eq:dimredaction1}) with $10$ replaced by $6$ and $n=2$ and we can
use that action to compute their Wald entropy using the Iyer-Wald prescription
\cite{Wald:1993nt,Iyer:1994ys}. 

The direct application of this prescription to Eq.~(\ref{eq:dimredaction1})
with $10$ replaced by a general dimension $D$ yields the following
string-frame entropy formula for $d=(D-n)$-dimensional black holes:
 
\begin{subequations}
  \begin{align}
  \label{eq:entropyd4}
S
  & =
  -2\pi\int_\Sigma d^{d-2}x\sqrt{|h|}
  \frac{\partial \mathcal{L}}{\partial R_{abcd}}
    \epsilon_{ab}\epsilon_{cd}\, ,
    \\
    & \nonumber \\
\label{eq:entropyformulastringframe}
      \frac{\partial\mathcal{L}}{\partial R_{abcd}}
    & =
      \frac{e^{-2(\phi-\phi_{\infty})}}{16\pi G_{N}^{(d)}}
      \left\{ g^{ab,\, cd}
      -\frac{\alpha'}{8} \left[H^{(0)\, abg}
        \left(\omega_{g}{}^{cd}-H^{(0)}_{g}{}^{cd}\right)
      \right.\right.
\nonumber  \\
    & \nonumber \\
    & \hspace{.5cm}
      \left. \left.
        -2R_{(-)}^{(0)\, abcd}
      +K^{(-)\,i\,[a|c}K^{(-)\,i\,|b]d} +K^{(+)\,j\,ab}K^{(+)\,j\,cd}
      \right]
      \right\}\, ,
  \end{align}
\end{subequations}

\noindent
where $|h|$ is the absolute value of the determinant of the metric induced
over the event horizon, $g^{ab,cd}= \tfrac{1}{2}(g^{ac}g^{bd}-g^{ad}g^{bc})$,
$\epsilon^{ab}$ is the event horizon's binormal normalized so that
$\epsilon_{ab}\epsilon^{ab}=-2$ and $R_{abcd}$ is the Riemann tensor.

This manifestly O$(n,n)$-invariant entropy formula reduces, for $n=1$ to the
formula found in Ref.~\cite{Elgood:2020xwu} and which has been used to compute
the Wald entropy of the $\alpha'$-corrected Reissner-Nordstr\"om black hole of
Ref.~\cite{Cano:2019ycn} and of the $\alpha'$-corrected heterotic version of
the Strominger-Vafa black hole of Ref.~\cite{Cano:2018qev}.

%%%%%%%%%%%%%%%%%%%%%%%%%%%%%%%%%%%%%%%%%%%%%%%%%%%%%%%%%%%%%%%%%%%%%%
%%%%%%%%%%%%%%%%%%%%%%%%%%%%%%%%%%%%%%%%%%%%%%%%%%%%%%%%%%%%%%%%%%%%%%
%%%%%%%%%%%%%%%%%%%%%%%%%%%%%%%%%%%%%%%%%%%%%%%%%%%%%%%%%%%%%%%%%%%%%%
%%%%%%%%%%%%%%%%%%%%%%%%%%%%%%%%%%%%%%%%%%%%%%%%%%%%%%%%%%%%%%%%%%%%%%
\subsection{The Wald entropy of the $\alpha'$-corrected $4d$ 4-charge black
  holes}
\label{sec-entropy4dbhs}
%%%%%%%%%%%%%%%%%%%%%%%%%%%%%%%%%%%%%%%%%%%%%%%%%%%%%%%%%%%%%%%%%%%%%%
%%%%%%%%%%%%%%%%%%%%%%%%%%%%%%%%%%%%%%%%%%%%%%%%%%%%%%%%%%%%%%%%%%%%%%
%%%%%%%%%%%%%%%%%%%%%%%%%%%%%%%%%%%%%%%%%%%%%%%%%%%%%%%%%%%%%%%%%%%%%%
%%%%%%%%%%%%%%%%%%%%%%%%%%%%%%%%%%%%%%%%%%%%%%%%%%%%%%%%%%%%%%%%%%%%%%

The $\alpha'$-corrected $4d$ 4-charge black-hole solutions correspond to the
following 10-dimensional solutions of the action Eq.~(\ref{heterotic}):

\begin{subequations}
  \begin{align}
\label{eq:metric}
ds^{2}
& = 
\frac{2}{\mathcal{Z}_{-}}du
\left[dv-\tfrac{1}{2}\mathcal{Z}_{+} du\right]
-\mathcal{Z}_{0} d\sigma^{2}
-dy^{i}dy^{i}\, ,
\,\,\,\,\,
i,j=1,2,3,4\, ,
\\
& \nonumber \\
\label{eq:H}
H 
& =  
d\mathcal{Z}^{-1}_{-}\wedge du \wedge dv
+\star_{(4)}d\mathcal{Z}_{0}\, ,  
\\
&  \nonumber \\
e^{-2{\phi}}
& = 
e^{-2{\phi}_{\infty}}\frac{\mathcal{Z}_{-}}{\mathcal{Z}_{0}}\, ,
  \end{align}
\end{subequations}

\noindent
where $d\sigma^{2}$ is the Gibbons-Hawking metric

\begin{equation}
\label{eq:GHmetric}
d\sigma^{2}= \mathcal{H}^{-1}(d\eta+\chi)^{2}+\mathcal{H}dx^{x}dx^{x}\, ,
\,\,\,\,\,
x,y,z=1,2,3\, ,
\,\,\,\,\,
d\mathcal{H}
=
\star_{(3)}d\chi\, ,
\end{equation}

\noindent
where $\star_{(3)}$ denotes the Hodge dual in $\mathbb{E}^{3}$. This last
equation implies that $\mathcal{H}$ is harmonic in $\mathbb{E}^{3}$. An
appropriate choice of harmonic function $\mathcal{H}$ for single,
spherically-symmetric, asymptotically flat black holes is

\begin{equation}
  \label{eq:exp_H}
\mathcal{H}
= 
1+\frac{\tilde{q}}{r}\, , 
\end{equation}

\noindent
and, then, the Gibbons-Hawking metric is that of a Kaluza-Klein monopole,
which, in spherical coordinates, takes the local form

\begin{equation}
d\sigma^{2}
=  
\mathcal{H}^{-1}(d\eta+\tilde{q}\cos{\theta} d\phi)^{2}
+\mathcal{H}\left(dr^{2}+r^{2}d\Omega^{2}_{(2)}\right)\, ,
\end{equation}

\noindent
where

\begin{equation}
d\Omega^{2}_{(2)}
=
d\theta^{2}+\sin^{2}\theta d\phi^{2}\, ,
\end{equation}

\noindent
and where $\eta$ parametrizes a circle of radius $R$. The Kaluza-Klein charge
$\tilde{q}$ has to be quantized according to

\begin{equation}
\label{quant}
\tilde{q} =\frac{W R}{2}\, , \quad W=1,2,\ldots
\end{equation}

\noindent
in order to avoid Dirac-Misner strings.

For this choice of $\mathcal{H}$ and to describe single,
spherically-symmetric, asymptotically-flat black holes, the functions
$\mathcal{Z}_{+},\mathcal{Z}_{-},\mathcal{Z}_{0},\mathcal{H}$ must take the
explicit form

\begin{subequations}
  \begin{align}
    \label{eq:exp_Z+}
    \mathcal{Z}_{+} 
    & = 
      1+\frac{\tilde{q}_{+}}{r}
      +\frac{\alpha'\tilde{q}_{+} }{2\tilde{q}\tilde{q}_{0}}
      \frac{r^{2}+r(\tilde{q}_{0}+\tilde{q}_{-}+\tilde{q})
      +\tilde{q}\tilde{q}_{0}
      +\tilde{q}\tilde{q}_{-}+\tilde{q}_{0}\tilde{q}_{-}}
      {(r-\tilde{q})(r+\tilde{q}_{0})(r+\tilde{q}_{-})}
      +\mathcal{O}(\alpha'{}^{2})\, ,
    \\
    & \nonumber\\
    \label{eq:exp_Z-}
    \mathcal{Z}_{-} 
    & = 
      1+\frac{\tilde{q}_{-}}{r}+\mathcal{O}(\alpha'{}^{2})\, ,
    \\
    & \nonumber\\
    \label{eq:exp_Z0}
    \mathcal{Z}_{0} 
    & = 
      1+\frac{\tilde{q}_{0}}{r}
      -\alpha'\left[
      \frac{(r+\tilde{q})(r+2\tilde{q}_{0})+\tilde{q}_{0}^{2}}
      {4\tilde{q}(r+\tilde{q})(r+\tilde{q}_{0})^{2}}
+      \frac{(r+\tilde{q})(r+2\tilde{q})+\tilde{q}^{2}}
      {4\tilde{q}(r+\tilde{q})^{3}}
      \right]
      +\mathcal{O}(\alpha'{}^{2})\, .
  \end{align}
\end{subequations}

As we discussed at the beginning of this section, we can compactify trivially
this solution in the T$^{4}$ parametrized by the coordinates
$y^{1},\cdots,y^{4}$. Relabeling $u=k_{\infty}z$, $\eta = \ell_{\infty}w$,
with $\ell_{\infty}\equiv R/\ell_{s}$\footnote{We normalize the periods of the
  compact coordinates that parametrize the internal circles, $z$ and $w$, to
  $2\pi\ell_{s}$. The information about the size of these circles is carried
  by the internal metric and the corresponding 4-dimensional moduli, which are
  dimensionless. Thus $G_{zz\,\infty}=k^{2}_{\infty}= (R_{z}/\ell_{s})^{2}$
  and $G_{ww\,\infty}=\ell^{2}_{\infty}= (R/\ell_{s})^{2}$.} and $v=t$, the
6-dimensional solution can be conveniently written in the following form:

\begin{subequations}
  \begin{align}
\label{eq:metric6}
d\hat{s}^{2}
    & =
          \frac{1}{\mathcal{Z}_{+}\mathcal{Z}_{-}}dt^{2}
-\mathcal{Z}_{0}\mathcal{H}\left(dr^{2}+r^{2}d\Omega^{2}_{(2)}\right)
    -k_{\infty}^{2}\frac{\mathcal{Z}_{+}}{\mathcal{Z}_{-}}
    \left(dz-    \frac{1}{k_{\infty}\mathcal{Z}_{+}}dt\right)^{2}
      \nonumber \\
    & \nonumber \\
    & \hspace{.5cm}
      -\ell_{\infty}^{2}\frac{\mathcal{Z}_{0}}{\mathcal{H}}
      \left(dw+\tilde{q}/\ell_{\infty}\cos{\theta} d\phi\right)^{2}\, ,
\\
&  \nonumber \\
\label{eq:H6}
\hat{H} 
& = 
d\left(-\frac{k_{\infty}}{\mathcal{Z}_{-}}  dt\right) \wedge dz 
+\ell_{\infty}r^{2}\mathcal{Z}_{0}'\,d\cos{\theta}\wedge d\phi\wedge dw \, ,  
\\
& \nonumber \\
e^{-2(\hat{\phi}-\hat{\phi}_{\infty})}
& =
\frac{\mathcal{Z}_{-}}{\mathcal{Z}_{0}}\, ,
  \end{align}
\end{subequations}

\noindent
where a prime indicates a derivative with respect to the radial coordinate
$r$. In this form we can immediately identify the KK scalars and vector fields

\begin{equation}
  G
  =
  \left(
    \begin{array}{cc}
      k_{\infty}^{2}\mathcal{Z}_{+}/\mathcal{Z}_{-} & 0 \\
                                                            & \\
      0 & \ell_{\infty}^{2}\mathcal{Z}_{0}/\mathcal{H} \\
     \end{array}
   \right)\, ,
   \hspace{1cm}
   \left(A^{m}{}_{\mu}dx^{\mu}\right)
   =
   \left(
     \begin{array}{c}
       -dt/(k_{\infty}\mathcal{Z}_{+}) \\
       \\
\tilde{q}/\ell_{\infty}\cos{\theta}d\phi \\       
     \end{array}
\right)\, ,
\end{equation}

\noindent
and the 4-dimensional string-frame metric

\begin{equation}
ds^{2}
=
          \frac{1}{\mathcal{Z}_{+}\mathcal{Z}_{-}}dt^{2}
-\mathcal{Z}_{0}\mathcal{H}\left(dr^{2}+r^{2}d\Omega^{2}_{(2)}\right)\, .
\end{equation}

The 4-dimensional dilaton field is

\begin{equation}
  e^{-2(\phi-\phi_{\infty})}
  =
  \sqrt{\frac{\mathcal{Z}_{+}\mathcal{Z}_{-}}{\mathcal{Z}_{0}\mathcal{H}}}\, ,
\end{equation}

\noindent
where

\begin{equation}
e^{-2\phi_{\infty}} =  e^{-2\hat{\phi}_{\infty}}k_{\infty}\ell_{\infty}\, ,
\end{equation}

\noindent
and the (modified) Einstein-frame metric is given by

\begin{equation}
  \begin{aligned}
    ds_{E}^{2}
    & =
    e^{2(\phi-\phi_{\infty})}ds^{2}
    = e^{-2U}dt^{2} -e^{2U}d\vec{x}^{\,2}\, ,
    \\
    & \\
    e^{2U} & =
    (\mathcal{Z}_{+}\mathcal{Z}_{-}\mathcal{Z}_{0}\mathcal{H})^{1/2}\, .
  \end{aligned}
\end{equation}

Let us now consider the Kalb-Ramond field strength and its
decomposition. First, we choose the following Vielbein basis

\begin{equation}
  \begin{array}{rclrcl}
    \hat{e}^{\,0} & = & {\displaystyle\frac{1}{\sqrt{\mathcal{Z}_{+}\mathcal{Z}_{-}}}}dt\, ,
    &    
  \hat{e}^{\,1} & = &  \sqrt{\mathcal{Z}_{0}\mathcal{H}}dr\, ,
    \\
    & & & & & \\
  \hat{e}^{\,2} & = &  \sqrt{\mathcal{Z}_{0}\mathcal{H}}rd\theta\, ,
    & 
      \hat{e}^{3} & = &  \sqrt{\mathcal{Z}_{0}\mathcal{H}} r\sin{\theta}d\phi\, ,
                        \\
                      & & & & & \\
    \hat{e}^{4} & = &  k_{\infty}\sqrt{\mathcal{Z}_{+}/\mathcal{Z}_{-}}
                      \left(dz-
                      {\displaystyle\frac{1}{k_{\infty}\mathcal{Z}_{+}}}dt\right)\, ,
                      \hspace{.5cm}
                      &
    \hat{e}^{5} & = &
\ell_{\infty}\sqrt{\mathcal{Z}_{0}/\mathcal{H}}
      \left(dw+\tilde{q}/\ell_{\infty}\cos{\theta} d\phi\right)\, ,    
  \end{array}
\end{equation}

\noindent
in terms of which, the non-vanishing components of $\hat{H}$ are

\begin{equation}
  \hat{H}_{104}
  =
  \frac{1}{\sqrt{\mathcal{Z}_{0}\mathcal{H}}}\left(\log{\mathcal{Z}_{-}}\right)'\,,
\hspace{1cm}
  \hat{H}_{235}
  =
  \frac{1}{\sqrt{\mathcal{Z}_{0}\mathcal{H}}}\left(\log{\mathcal{Z}_{0}}\right)'\,.
\end{equation}

This implies that the 4-dimensional Kalb-Ramond field strength vanishes
identically and there are two non-vanishing 4-dimensional winding vector field
strengths plus, perhaps, scalars. Computing the 4-dimensional Kalb-Ramond
2-form $B^{(1)}{}_{\mu\nu}$, the winding vectors $B^{(1)}{}_{m\,\mu}$ and the
scalars $B^{(1)}{}_{mn}$ is very complicated because of the $\alpha'$
corrections in their definitions. Fortunately for us, only their zeroth-order
values contribute to the entropy formula. At this order 

\begin{equation}
\hat{H}^{(0)} 
= 
d\left(-\frac{k_{\infty}}{\mathcal{Z}_{-}}  dt\right) \wedge dz 
+\left(-\ell_{\infty}\tilde{q}_{0}\,\cos{\theta}d\phi\right)\wedge dw \, ,  
\end{equation}

\noindent
from which we read the only non-vanishing fields descending from the
Kalb-Ramond field:

\begin{equation}
  \left(B^{(0)}{}_{m\, \mu}dx^{\mu}\right)
  =
  \left(
    \begin{array}{c}
      -k_{\infty}/\mathcal{Z}_{-}  dt \\ \\
      -\ell_{\infty}\tilde{q}_{0}\,\cos{\theta}d\phi \\
    \end{array}
    \right)\, .
\end{equation}

This allows us to compute the nonvanishing components of the 2-forms
$K^{(0)\,i}_{(\pm)}$:

\begin{equation}
  \begin{aligned}
    K^{(0)\,4\,01}_{(\pm)}
    & =
    \frac{1}{\sqrt{\mathcal{Z}_{0}\mathcal{H}}}
    \left\{
      \left(\log{\mathcal{Z}_{+}}\right)'\pm
      \left(\log{\mathcal{Z}_{-}}\right)'
     \right\}\, ,
    \\
    & \\
    K^{(0)\,5\,23}_{(\pm)}
    & =
    \frac{1}{r^{2}(\mathcal{Z}_{0}\mathcal{H})^{3/2}}
    \left(\tilde{q}\mathcal{Z}_{0}
    \mp \tilde{q}_{0}\mathcal{H}
    \right)\, .
  \end{aligned}
\end{equation}

Taking into account the vanishing of the 4-dimensional Kalb-Ramond field
strength, the entropy formula takes the simple form

\begin{equation}
  \begin{aligned}
  \label{eq:entropy1}
S
  & =
  -\frac{1}{8 G_{N}^{(4)}}
\int_{\Sigma} d^{2}x\sqrt{|h|}e^{-2(\phi-\phi_{\infty})}
  \epsilon_{ab}\epsilon_{cd}
      \left\{ g^{ab,\, cd}
\right.
\\
    & \\
    & \hspace{.5cm}
 \left.
      -\frac{\alpha'}{8} \left[-2R^{(0)\, abcd}
      +K^{(-)\,i\,ac}K^{(-)\,i\,bd} +K^{(+)\,j\,ab}K^{(+)\,j\,cd}
      \right]
      \right\}
      \\
      & \\
      & =
  \frac{1}{4 G_{N}^{(4)}}
\int_{\Sigma} d^{2}x\sqrt{|h|}e^{-2(\phi-\phi_{\infty})}
\left\{ 1
  +\frac{\alpha'}{4} \left[-2R^{(0)\, 0101}
      +\left(K^{(-)\,4\,01}\right)^{2} +\left(K^{(+)\,4\,01}\right)^{2}
      \right]
      \right\}
      \\
      & \\
      & =
  \frac{1}{4G_{N}^{(4)}}
\left\{ A_{\mathcal{H}}
  +2\pi\alpha'\lim_{r\rightarrow 0}
  e^{2U}r^{2}  \left[
    -\sqrt{\frac{\mathcal{Z}_{+}\mathcal{Z}_{-}}{\mathcal{Z}_{0}\mathcal{H}}}
      \left[\frac{1}{\sqrt{\mathcal{Z}_{0}\mathcal{H}}}
        \left(\frac{1}{\sqrt{\mathcal{Z}_{+}\mathcal{Z}_{-}}}\right)'\right]'
    \right.
        \right.
    \\
    & \\
    & \hspace{.5cm}
    \left.
          \left.
    +\frac{1}{\mathcal{Z}_{0}\mathcal{H}}
    \left[
      \left(\frac{\mathcal{Z}_{+}'}{\mathcal{Z}_{+}}\right)^{2}
      +\left(\frac{\mathcal{Z}_{-}'}{\mathcal{Z}_{-}}\right)^{2}
     \right]
      \right]
      \right\}
%       \\
%       & \\
%       & =
%   \frac{1}{4G_{N}^{(4)}}
% \left\{ A_{\mathcal{H}}
%   +2\pi\alpha' \sqrt{\tilde{q}_{+}\tilde{q}_{-}\tilde{q}_{0}\tilde{q}} \left[ 
%     -\frac{1}{\tilde{q}_{0}\tilde{q}}
%   +\frac{2}{\tilde{q}_{0}\tilde{q}}
%       \right]
%       \right\}
      \\
      & \\
      & =
  \frac{ A_{\mathcal{H}}}{4G_{N}^{(4)}}
\left\{1
     +\frac{\alpha'}{2\tilde{q}_{0}\tilde{q}}
      \right\}
    \end{aligned}
\end{equation}

\noindent
where we have assumed that all the charges $\tilde{q}_{i}$ are different from
zero\footnote{We have assumed the positivity of all the charge parameters
  because, as a general rule, if these parameters have negative values, there
  are naked singularities. However, the $\alpha'$ corrections can sometimes
  eliminate these singularities, as shown in Ref.~\cite{Cano:2018aod}. On the
  other hand, the signs of the charges are not completely determined by those
  parameters, but by additional coefficients ($\pm 1$) that appear multiplying
  them in the 4-dimensional gauge fields. These coefficients have the values
  corresponding to supersymmetric black holes in the example that we are
  considering.} and where

\begin{equation}
  A_{\mathcal{H}}
  =
  4\pi\sqrt{\tilde{q}_{+}\tilde{q}_{-}\tilde{q}_{0}\tilde{q}}\, ,
\end{equation}

\noindent
is the area of the horizon, which does not receive any corrections to first
order in $\alpha'$.

The charges $\tilde{q}_{i}$ are related to the numbers of solitonic 5-branes
$N$, units of momentum $n$, winding number $w$ and KK charge $W$ by
\cite{Cano:2018brq}\footnote{Here we are using the notation of
  Ref.~\cite{Faedo:2019xii}.}

\begin{equation}
\tilde{q}_{0}=\frac{\alpha'}{2 R} N\, ,
\qquad 
\tilde{q}_{+} = \frac{ \alpha'{}^{2} g_{s}^{2}}{2 R R_{z}^{2}} n\, ,
\qquad 
\tilde{q}_{-} = \frac{\alpha' g_{s}^{2}}{2 R} w\, ,
\qquad 
q = \frac{W R}{2}\, ,
\end{equation}

\noindent
and, using the value of the 10-dimensional Newton constant in
Eq.~(\ref{eq:d10newtonconstant}) and its relation with the 4-dimensional one
Eq.~(\ref{eq:relationsbetweenconstants}) with
$V_{6}=(2\pi)^{6}\alpha'{}^{2}RR_{z}$) we get

\begin{equation}
  S
  =
  2\pi \sqrt{NnWw}\left\{1 +\frac{2}{NW}\right\}\, .
\end{equation}

Finally, using the relation between the numbers of objects and the asymptotic
charges found in Ref.~\cite{Faedo:2019xii}

\begin{subequations}
  \begin{align}
    \mathcal{Q}_{+}
    & =
      n\left(1+\frac{2}{NW}\right)\, ,
    \\
    & \nonumber \\
    \mathcal{Q}_{-}
    & =
      w\, ,
    \\
    & \nonumber \\
    \mathcal{Q}_{0}
    & =
      N\left(1-\frac{2}{NW}\right)\, ,
    \\
    & \nonumber \\
    \mathcal{Q}
    & =
      W\, ,
  \end{align}
\end{subequations}

\noindent
we find that

\begin{equation}
S = 2\pi \sqrt{\mathcal{Q}_{+}\mathcal{Q}_{-}(\mathcal{Q}_{0}\mathcal{Q}+4)}\, .  
\end{equation}

This is the entropy obtained by microstate counting in
Ref.~\cite{Kutasov:1998zh}.

The entropy formula we have derived can, evidently, be applied to other, more
general, solutions, such as 4-charge extremal non-supersymmetric and
non-extremal black holes. The non-supersymmetric extremal ones can be easily
obtained from the supersymmetric ones by flipping the sign of a charge
\cite{Ortin:1996bz}, but a computation of the $\alpha'$ corrections to their
entropy (which at lowest order in $\alpha'$ is identical to that of the
supersymmetric ones) requires a computation of the $\alpha'$ corrections to
the solutions themselves. This is, nevertheless, an interesting problem whose
result should be compared with the result obtained in
Ref.~\cite{LopesCardoso:1999fsj} by a different method.

To conclude this section it is, perhaps, worth stressing that the result we
have obtained differs from the result obtained in Ref.~\cite{Cano:2018brq} and
it is also worth explaining why.

There is a potential problem with all the results in this field of work: the
dangers of confirmation bias. It is easy to conclude that, if the microscopic
entropy is reproduced in a certain macroscopic calculation, then this
calculation is correct. This temptation is strengthened by the fact that it is
very difficult to make additional checks on the calculation, specially when
one considers extremal black holes with zero temperature only.

The result of Ref.~\cite{Cano:2018brq}, obtained using the Iyer-Wald
prescription directly in 10 dimensions agrees with the microscopic calculation
if it is interpreted as an approximate result containing the zeroth- and
first-order terms in an expansion in $\alpha'$ of the microscopic result.
However, further research carried out in Ref.~\cite{Faedo:2019xii} showed that
the procedure followed in Ref.~\cite{Cano:2018brq} had several problems. In
particular, the dependence of the action on the curvature in 10 or in 10-d
dimensions is different and, applying the Iyer-Wald formula to it gives
different results. The dimensional reduction of the action to $d=10-n$
dimensions carried out in this paper was needed to obtain the correct
dependence on the curvature. The $n=1$ case was dealt with in
Ref.~\cite{Elgood:2020xwu} for the same reasons but at least $n=2$ was needed
to consider 4-charge black holes.

The difference between the result of Ref.~\cite{Cano:2018brq} and the result
obtained applying the Iyer-Wald prescription to the dimensionally-reduced
action is very small: the coefficient of one term changes from 1 to 2. The
effects of this change are, however, important:

\begin{enumerate}
\item The entropy formula obtained in Ref.~\cite{Elgood:2020xwu} with this
  coefficient gives an entropy which satisfies the first law of thermodynamics
  in the case of a non-extremal, $\alpha'$-corrected Reissner-Nordstr\"om
  black hole. This provides the additional check one needs to make sure the
  result is correct, avoiding any confirmation bias. There is no microscopic
  entropy calculation for this particular black hole, but that is a completely
  different problem.
  
\item The entropy formula obtained in Ref.~\cite{Elgood:2020xwu}, when applied
  to the $\alpha'$-corrected Strominger-Vafa black hole, gives a result for
  the entropy different from the one obtained in our paper
  \cite{Cano:2018qev}, which is analogous to Ref.~\cite{Cano:2018brq} but
  deals with 3-charge black holes. However, this entropy now coincides
  identically (not as an approximation) with the microscopic entropy.

\item Now the same happens in the case that we have considered in this paper:
  the entropy is different from the one obtained in Ref.~\cite{Cano:2018brq},
  which was interpreted as an approximation, but now it gives the exact
  result.
\end{enumerate}

Using the dimensionally-reduced action has solved one of the problems that
arise in the application of Iyer-Wald prescription to (heterotic) stringy
black holes. The rest of the problems discussed in the introduction to
Ref.~\cite{Elgood:2020xwu} still remain, as we are going to discuss in the
last section.

%%%%%%%%%%%%%%%%%%%%%%%%%%%%%%%%%%%%%%%%%%%%%%%%%%%%%%%%%%%%%%%%%%%%%% 
%%%%%%%%%%%%%%%%%%%%%%%%%%%%%%%%%%%%%%%%%%%%%%%%%%%%%%%%%%%%%%%%%%%%%% 
%%%%%%%%%%%%%%%%%%%%%%%%%%%%%%%%%%%%%%%%%%%%%%%%%%%%%%%%%%%%%%%%%%%%%%
%%%%%%%%%%%%%%%%%%%%%%%%%%%%%%%%%%%%%%%%%%%%%%%%%%%%%%%%%%%%%%%%%%%%%%
\section{Conclusions}
\label{sec-conclusions}
%%%%%%%%%%%%%%%%%%%%%%%%%%%%%%%%%%%%%%%%%%%%%%%%%%%%%%%%%%%%%%%%%%%%%%
%%%%%%%%%%%%%%%%%%%%%%%%%%%%%%%%%%%%%%%%%%%%%%%%%%%%%%%%%%%%%%%%%%%%%%
%%%%%%%%%%%%%%%%%%%%%%%%%%%%%%%%%%%%%%%%%%%%%%%%%%%%%%%%%%%%%%%%%%%%%%
%%%%%%%%%%%%%%%%%%%%%%%%%%%%%%%%%%%%%%%%%%%%%%%%%%%%%%%%%%%%%%%%%%%%%%

In this paper we have shown that the bosonic sector of the complete Heterotic
Superstring effective action compactified on T$^{n}$ is O$(n,n)$ invariant to
first-order in $\alpha'$.\footnote{A similar result using the effective action
  of Ref.~\cite{Metsaev:1987zx} and without Yang-Mills fields has been
  recently published in Ref.~\cite{Eloy:2020dko}. Furthermore, it has been
  argued in Ref.~\cite{Hohm:2014sxa} that O$(n,n)$ is present at all orders in
  $\alpha'$.} Local supersymmetry (which is preserved by the $\alpha'$
corrections) ensures the O$(n,n)$ invariance of the whole theory at first
order in $\alpha'$. The $(10-n)$-dimensional action is not really suitable for
the derivation of the $(10-n)$-dimensional equations of motion because it is
not easy to apply the Bergshoeff-de Roo lemma \cite{Bergshoeff:1989de} to
it. Nevertheless, one can always work in 10 dimensions, where it is easy to
apply it, reducing afterwards the solutions to $(10-n)$ using the formulae
obtained in this paper.\footnote{One can replace 10 by $D$ in this discussion,
  as explained at the beginning of Section~\ref{sec-entropyformula}.}
Furthermore, the action can be used to obtain a Wald entropy formula which we
have used to compute the entropy of the most basic stringy 4-dimensional,
4-charge black holes. In order to apply it to more general black holes,
non-extremal, non-supersymmetric, with more vector and scalar fields or with
angular momentum one first needs to find their $\alpha'$ corrections, which
can be a non-trivial, albeit highly interesting, task.

Our results leave, however, some important questions unanswered:

\begin{itemize}

\item If the vector fields are Abelian, does one recover O$(n,n+n_{V})$
  invariance in the presence of all the $\alpha'$ corrections (that is: adding
  the torsionful spin connection terms)? At first sight the answer should be
  yes: the situation is not qualitatively different from having a number of
  gauge fields $n_{A}$ of which are Abelian and $n_{N}$ of which are
  non-Abelian, where one should have O$(n,n+n_{A})$ invariance. Nevertheless,
  it would be convenient to rewrite this case in a manifestly
  O$(n,n+n_{V})$-invariant form.

\item It is tempting to conjecture that this invariance will be maintained to
  all orders in $\alpha'$. However, since this invariance depends on the
  conspiracy between quite different terms, and we have not found the
  systematics behind the invariance, more work would be necessary in order to
  justify such a conjecture.

\item In the 4-dimensional case, is S~duality preserved (as expected)
  \cite{Font:1990gx}? How is it realized?

  Work in these directions is already in progress \cite{kn:EMMO}.

\item Why is the Iyer-Wald prescription so successful in this setting? As we
  have explained in the introduction to Ref.~\cite{Elgood:2020xwu}, the
  Iyer-Wald prescription is based on two conditions: the invariance of the
  action under diffeomorphisms and the tensorial character of all the fields
  in the theory. While the first condition is met by the 10- and
  $(10-n)$-dimensional actions Eqs.~(\ref{heterotic}) and
  (\ref{eq:dimredaction1}) to the order in $\alpha'$ we are considering
  here,\footnote{There seems to be some confusion in the literature concerning
    this point. Sometimes the presence of Lorentz-Chern-Simons terms in the
    Kalb-Ramond field strength is associated to lack of invariance of the
    action under diffeomorphisms. However, the action Eqs.~(\ref{heterotic})
    is, by construction, invariant under Yang-Mills, local-Lorentz, general
    coordinate and local-supersymmetry transformations order by order in
    $\alpha'$ \cite{Bergshoeff:1989de}. If the Kalb-Ramond 2-form is dualized
    into a 6-form, the Chern-Simons 3-forms disappear from the field strength
    but reappear in a Chern-Simons 10-form \cite{Bergshoeff:1990ax}. That dual
    action is then invariant up to a total derivative which can be dealt with
    by the procedure considered in Ref.~\cite{Tachikawa:2006sz}. This
    reference does not deal with the gauge freedoms of the fields of the
    theory, though. In any case, one should not confuse the actions containing
    Chern-Simons 10- and the 3-forms, because the latter (ours) is exactly
    invariant under diffeomorphisms, which is the reason why we do not need
    the results of Ref.~\cite{Tachikawa:2006sz}.} the second is not because
  Yang-Mills fields and Vielbeins have gauge freedoms, a fact recognized and
  partially dealt with in
  Refs.~\cite{Compere:2007vx,Jacobson:2015uqa,Prabhu:2015vua,Elgood:2020svt,Aneesh:2020fcr}. The
  case of a Kalb-Ramond field with Nicolai-Townsend gauge transformations
  (which is the case that arises in this formulation of the Heterotic
  superstring effective action) has not yet been fully studied.

  This problem with the applicability of the Iyer-Wald prescription to the
  Heterotic Superstring effective action results, as a matter of fact, in
  entropy formulae which are manifestly not invariant under local Lorentz
  transformations.\footnote{A more rigorous derivation, performed in
    Ref.~\cite{Edelstein:2018ewc} which takes into account these
    transformations shows the presence of an additional term not captured by
    the Iyer-Wald prescription that may restore the (expected) local Lorentz
    invariance of the Wald entropy formula. This term may not contribute in
    many relevant cases but it remains to be explained why and when. On the
    other hand, a proof of the first law of black hole mechanics for the
    theory at hands may have to be found in order to identify unambiguously
    the entropy.} And, yet, the results we have obtained by using this formula
  here and in Refs.~\cite{Cano:2019ycn,Elgood:2020xwu} seem to be quite
  satisfactory, which we interpret as an indication that this formula captures
  the essential terms that contribute to the entropy.

  It is clear that much more work is necessary to find an unambiguous,
  Lorentz-invariant, entropy formula for the Heterotic Superstring effective
  action. Hopefully, such a formula must reduce under certain general
  conditions to the entropy formula we have obtained here, but only when this
  formula is found it will be possible to make a fully trustable macroscopic
  calculation of the black hole entropy that we can compare with the
  microscopic one.

  Work in this direction is also well under way \cite{kn:EMPO}.

\end{itemize}

%%%%%%%%%%%%%%%%%%%%%%%%%%%%%%%%%%%%%%%%%%%%%%%%%%%%%%%%%%%%%%%%%%%%%% 
%%%%%%%%%%%%%%%%%%%%%%%%%%%%%%%%%%%%%%%%%%%%%%%%%%%%%%%%%%%%%%%%%%%%%%
%%%%%%%%%%%%%%%%%%%%%%%%%%%%%%%%%%%%%%%%%%%%%%%%%%%%%%%%%%%%%%%%%%%%%%
%%%%%%%%%%%%%%%%%%%%%%%%%%%%%%%%%%%%%%%%%%%%%%%%%%%%%%%%%%%%%%%%%%%%%%
\section*{Acknowledgments}
%%%%%%%%%%%%%%%%%%%%%%%%%%%%%%%%%%%%%%%%%%%%%%%%%%%%%%%%%%%%%%%%%%%%%%
%%%%%%%%%%%%%%%%%%%%%%%%%%%%%%%%%%%%%%%%%%%%%%%%%%%%%%%%%%%%%%%%%%%%%%
%%%%%%%%%%%%%%%%%%%%%%%%%%%%%%%%%%%%%%%%%%%%%%%%%%%%%%%%%%%%%%%%%%%%%%
%%%%%%%%%%%%%%%%%%%%%%%%%%%%%%%%%%%%%%%%%%%%%%%%%%%%%%%%%%%%%%%%%%%%%%

The author would like to thank P.~Meessen for useful conversations and
A.~Ruip\'erez for pointing to us an error in an equation. This work
has been supported in part by the MCIU, AEI, FEDER (UE) grant
PGC2018-095205-B-I00 and by the Spanish Research Agency (Agencia Estatal de
Investigaci\'on) through the grant IFT Centro de Excelencia Severo Ochoa
SEV-2016-0597.  TO wishes to thank M.M.~Fern\'andez for her permanent support.

%%%%%%%%%%%%%%%%%%%%%%%%%%%%%%%%%%%%%%%%%%%%%%%%%%%%%%%%%%%%%%%%%%%%%% 
%%%%%%%%%%%%%%%%%%%%%%%%%%%%%%%%%%%%%%%%%%%%%%%%%%%%%%%%%%%%%%%%%%%%%%
%%%%%%%%%%%%%%%%%%%%%%%%%%%%%%%%%%%%%%%%%%%%%%%%%%%%%%%%%%%%%%%%%%%%%%
%%%%%%%%%%%%%%%%%%%%%%%%%%%%%%%%%%%%%%%%%%%%%%%%%%%%%%%%%%%%%%%%%%%%%%
%%%%%%%%%%%%%%%%%%%%%%%%%%%%%%%%%%%%%%%%%%%%%%%%%%%%%%%%%%%%%%%%%%%%%%
%%%%%%%%%%%%%%%%%%%%%%%%%%%%%%%%%%%%%%%%%%%%%%%%%%%%%%%%%%%%%%%%%%%%%%
\appendix
%%%%%%%%%%%%%%%%%%%%%%%%%%%%%%%%%%%%%%%%%%%%%%%%%%%%%%%%%%%%%%%%%%%%%%
%%%%%%%%%%%%%%%%%%%%%%%%%%%%%%%%%%%%%%%%%%%%%%%%%%%%%%%%%%%%%%%%%%%%%%
%%%%%%%%%%%%%%%%%%%%%%%%%%%%%%%%%%%%%%%%%%%%%%%%%%%%%%%%%%%%%%%%%%%%%%
%%%%%%%%%%%%%%%%%%%%%%%%%%%%%%%%%%%%%%%%%%%%%%%%%%%%%%%%%%%%%%%%%%%%%%

%%%%%%%%%%%%%%%%%%%%%%%%%%%%%%%%%%%%%%%%%%%%%%%%%%%%%%%%%%%%%%%%%%%%%%
%%%%%%%%%%%%%%%%%%%%%%%%%%%%%%%%%%%%%%%%%%%%%%%%%%%%%%%%%%%%%%%%%%%%%%
%%%%%%%%%%%%%%%%%%%%%%%%%%%%%%%%%%%%%%%%%%%%%%%%%%%%%%%%%%%%%%%%%%%%%%
%%%%%%%%%%%%%%%%%%%%%%%%%%%%%%%%%%%%%%%%%%%%%%%%%%%%%%%%%%%%%%%%%%%%%%
\section{$\mathrm{O}(n,n)/(\mathrm{O}(n)\times \mathrm{O}(n))$ coset space}
\label{app-Onn}
%%%%%%%%%%%%%%%%%%%%%%%%%%%%%%%%%%%%%%%%%%%%%%%%%%%%%%%%%%%%%%%%%%%%%% 
%%%%%%%%%%%%%%%%%%%%%%%%%%%%%%%%%%%%%%%%%%%%%%%%%%%%%%%%%%%%%%%%%%%%%%
%%%%%%%%%%%%%%%%%%%%%%%%%%%%%%%%%%%%%%%%%%%%%%%%%%%%%%%%%%%%%%%%%%%%%%
%%%%%%%%%%%%%%%%%%%%%%%%%%%%%%%%%%%%%%%%%%%%%%%%%%%%%%%%%%%%%%%%%%%%%%

Since this coset space occurs repeatedly in the main body of this paper, we
describe it here in some detail. First, we define the $n\times n$ matrices
$E\equiv(e^{i}{}_{m})$, $G\equiv (G_{mn})$ and $B\equiv (B_{mn})$. Evidently
$E^{-1}=(e^{m}{}_{i})$ and $E^{\rm T}E= G$. With them we  construct the $2n\times 2n$
Vielbein $\mathcal{U}$ and its inverse $\mathcal{U}^{-1}$

\begin{subequations}
  \begin{align}
  \mathcal{U}
& =
-\tfrac{1}{\sqrt{2}}\!\left(
 \begin{array}{c@{\quad}c}
 E^{-1} & E^{-1} \\[-6pt]
 &\\
 E^{\rm T}+BE^{-1} & -E^{\rm T}+BE^{-1}
 \end{array}
\right)
  =
  -\tfrac{1}{\sqrt{2}}\!\left(
 \begin{array}{c@{\quad}c}
 e^{m}{}_{i} & -e^{mi} \\[-6pt]
 & \\
 -e_{mi} +B_{mn}e^{n}{}_{i} &  -e_{m}{}^{i} -B_{mn}e^{ni} \\
 \end{array}
    \right)\, ,
    \\
    & \nonumber \\
    \mathcal{U}^{-1}
    & =
\tfrac{1}{\sqrt{2}}\!\left(
 \begin{array}{c@{\quad}c}
 -E+E^{-1\, {\rm T}}B & -E^{-1\, {\rm T}} \\
 &\\[-6pt]
 -E-E^{-1\, {\rm T}}B & E^{-1\, {\rm T}} \\
 \end{array}
    \right)
    =
\tfrac{1}{\sqrt{2}}\!\left(
 \begin{array}{c@{\quad}c}
-e^{i}{}_{m}-e^{in}B_{nm} & e^{im}  \\
&\\[-6pt]
e_{im} -e_{i}{}^{n}B_{nm} & e_{i}{}^{m}  \\
 \end{array}
    \right)    
  \end{align}
\end{subequations}

\noindent
where we have used the metric $-\delta_{ij}$ to raise and lower SO$(n)$
indices $i,j$, consistently with our mostly minus convention for the
10-dimensional spacetime metric.

In terms of  the non-diagonal O$(n,n)$ metric
$\Omega=\Omega^{-1}=\left(
  \begin{array}{cc}
    0 & \mathbb{1}_{n\times n} \\
    \mathbb{1}_{n\times n} & 0 \\
  \end{array}
\right)$ we can define O$(n,n)$ transformations $\Lambda$ as those satisfying

\begin{equation}
 \Lambda^{\rm T} \Omega\Lambda  =  \Omega\, .  
\end{equation}

\noindent
Under a O$(n,n)$ transformation $\Lambda$ acting from the left, the Vielbein
$\mathcal{U}$ transforms as

\begin{equation}
  \Lambda \mathcal{U} = \mathcal{U}' R\, ,
  \hspace{1cm}
  R
  =
  \left(
    \begin{array}{cc}
      R_{(+)} & 0 \\
      0 & R_{(-)} \\
    \end{array}
    \right)\in {\rm O}(n)_{(+)}\times {\rm O}(n)_{(-)}\, .
\end{equation}

\noindent
Thus, using $\mathcal{U}$, we can construct a symmetric matrix $M$  

\begin{equation}
 \label{eq:Mmatrixdef}
 M
 \equiv
 \mathcal{U}\mathcal{U}^{\rm T}=
  \left(
    \begin{array}{cc}
      G^{-1} & -G^{-1}B \\
             & \\
      -B^{\rm T}G^{-1} &   G+B^{\rm T}G^{-1}B \\
    \end{array}
    \right)\, ,
\end{equation}

\noindent
which transforms under O$(n,n)$ as

\begin{equation}
M' = \Lambda M \Lambda^{\rm T}\, .  
\end{equation}

\noindent
The inverse of $M$ is given by

\begin{equation}
  M^{-1}
  =
  \mathcal{U}^{-1\, \rm T}\mathcal{U}^{-1}
  =
    \left(
    \begin{array}{cc}
      G +B^{\rm T}G^{-1}B & -B^{\rm T}G^{-1} \\
                                 & \\
      -G^{-1}B^{\rm T} & G^{-1} \\
    \end{array}
    \right)\, ,
\end{equation}

\noindent
and it is not difficult to check that

\begin{equation}
M^{-1}=\Omega M \Omega\, ,  
\end{equation}

\noindent
which implies that $M$ is a O$(n,n)$ matrix itself.

The left-invariant Maurer-Cartan 1-form is

\begin{equation}
  -\mathcal{U}^{-1}d\mathcal{U}
  =
  \left(
    \begin{array}{c@{\quad}c}
      A_{(+)}{}^{i}{}_{j} & P_{(+)}{}^{ij} \\
                          & \\
   P_{(-)\,ij} &  A_{(-)\, i}{}^{j} \\
    \end{array}
    \right)\, ,
\end{equation}

\noindent
where the the O$(n)_{(+)}$ connection $A_{(+)}{}^{i}{}_{j}$, the O$(n)_{(-)}$
connection $A_{(-)\,i}{}^{j}$ and the Vielbein $P_{(+)\,ij}=P_{(-)\,ji}$
($n^{2}$ degrees of freedom) are given by\footnote{We have made use of
  Eq.~(\ref{eq:ede=eedG}).}
 
\begin{subequations}
  \begin{align}
    A_{(\pm)}{}^{ij}
    & \equiv
      A^{ij} \mp\tfrac{1}{2}e^{im}e^{jn}dB_{mn}\, , \\
    & \nonumber \\
    P_{(\pm)}{}^{ij}
    & \equiv
      \tfrac{1}{2}e^{im}e^{jn}d(G_{mn}\mp B_{mn})\, ,
  \end{align}
\end{subequations}

\noindent
where, in its turn

\begin{equation}
    \label{eq:sonconnection}
A^{ij}\equiv  -e^{[i|\, m}de_{m}{}^{|j]}\, .
\end{equation}

\noindent
Observe that the Vielbein transforms under both O$(n)$ groups:

\begin{equation}
P'_{(\pm)} = R_{\pm}P_{(\pm)}R_{\mp}\, . 
\end{equation}

The Maurer-Cartan equations, obtained by taking the exterior derivative of the
left-invariant Maurer-Cartan 1-forms lead to the following identities:

\begin{subequations}
  \begin{align}
    \label{eq:MCE1}
    F_{(+)}{}^{ij}
    & =
      P_{(+)}{}^{ik}\wedge P_{(-)\,k}{}^{j}\, ,
    \\
    & \nonumber \\
    \label{eq:MCE2}
    F_{(-)\,ij}
    & =
      P_{(-)\,ik}\wedge P_{(+)}{}^{k}{}_{j}\, ,
    \\
    & \nonumber \\
    \mathfrak{D}_{(\pm\mp)}P_{(\pm)}{}^{ij}
    & =
      0\, ,
  \end{align}
\end{subequations}

\noindent
where

\begin{equation}
F_{(\pm)}{}^{ij}=  dA_{(\pm)}{}^{ij}+A_{(\pm)}{}^{ik}\wedge A_{(\pm)}{}^{kj}\, ,
\end{equation}

\noindent
are the curvatures of the two O$(n)$ connections and where the covariant
derivative $\mathfrak{D}_{(\mp\pm)}$ uses the $(\mp)$ connection on the first
O$(n)$ index and the $(\pm)$ connection on the second of $P_{(\pm)}{}^{ij}$,
that is:

\begin{equation}
  \mathfrak{D}_{(\pm\mp)}P_{(\pm)}{}^{ij}
   =
  dP_{(\pm)}{}^{ij}+A_{(\pm)}{}^{ik}\wedge P_{(\pm)}{}^{kj}
    +\wedge P_{(\pm)}{}^{ik}\wedge A_{(\mp)}{}^{kj}\, .
\end{equation}

From the Maurer-Cartan Eqs.~(\ref{eq:MCE1}) and (\ref{eq:MCE2}) we find

\begin{equation}
  \label{eq:FFPPPP}
  F_{(\pm)}{}^{ij}\cdot F_{(\pm)}{}^{ij}
  =
  2P_{(\pm)}^{(0)\,a}{}_{ki}P_{(\pm)}^{(0)}{}_{a\,li}
  P_{(\pm)}^{(0)\,b}{}_{ki}P_{(\pm)}^{(0)}{}_{b\,li}
  -2P_{(\pm)}^{(0)\,a}{}_{ki}P_{(\pm)}^{(0)}{}_{a\,lj}
  P_{(\pm)}^{(0)\,b}{}_{li}P_{(\pm)}^{(0)}{}_{b\,kj}\, ,
\end{equation}

\noindent
and the relation $P_{(\pm)\,ij}=P_{(\mp)\,ji}$ implies that

\begin{equation}
  F_{(+)}{}^{ij}\cdot F_{(+)}{}^{ij}
  =
    F_{(-)}{}^{ij}\cdot F_{(-)}{}^{ij}\, .
\end{equation}

The scalar kinetic term is given by

\begin{equation}
  \begin{aligned}
    P_{(+)\,\mu}{}^{ij}P_{(-)}{}^{\mu}{}_{ji}
    & =
    \tfrac{1}{4}G^{mn}G^{pq}
    \partial_{\mu}(G_{mp}+B_{mp})\partial^{\mu}(G_{nq}-B_{nq})
    \\
    & \\
    & = -\tfrac{1}{4}
    \partial_{\mu}G_{mn} \partial_{\mu}G^{mn}
    -\tfrac{1}{4}G^{mn}G^{pq}\partial_{\mu}B_{mp}\partial^{\mu}B_{nq}\, ,
  \end{aligned}
\end{equation}

\noindent
or by the equivalent expression

\begin{equation}
-\tfrac{1}{2}\mathrm{Tr}\left(\partial_{\mu} M \partial^{\mu} M^{-1}\right)\, .  
\end{equation}

Using the O$(n,n)$ vector

\begin{equation}
 \mathcal{F}
  \equiv
  \left(
    \begin{array}{c}
      F^{m} \\ G_{m} 
    \end{array}
    \right)\, ,
\end{equation}

\noindent
one can construct the following combinations that occur naturally in some of
the expressions in the main paper:

\begin{equation}
  \label{eq:Kpmdef}
  \mathcal{U}^{-1}\mathcal{F}
  =
  \tfrac{1}{\sqrt{2}}\left(
    \begin{array}{c}
     e^{i}{}_{m}\left(F^{m}+G^{mn}G^{(0)}{}_{n}-G^{mn}B_{np}F^{p}\right)
      \\
      \\
     e^{i}{}_{m}\left(F^{m}-G^{mn}G^{(0)}{}_{n}+G^{mn}B_{np}F^{p}\right)
    \end{array}
  \right)
  \equiv
-\tfrac{1}{\sqrt{2}}\left(
  \begin{array}{c}
   K_{(+)}{}^{i} \\ \\  K_{(-)}{}^{i}
  \end{array}
  \right)
  \, .
\end{equation}

Observe that $K_{(\pm)}{}^{i}$ only transform under O$(n)_{\pm}$ rotations,
respectively. We can construct O$(n,n)$ invariants by building O$(n)_{\pm}$
invariants. For instance:

\begin{equation}
  K_{(\pm)\,i} K_{(\pm)\,i}
  =
  \mathcal{F}^{\rm T}M^{-1}\mathcal{F}
  \pm
  2\mathcal{F}^{\rm T}\Omega\mathcal{F}\, ,
\end{equation}

\noindent
which is clearly consistent with 

\begin{equation}
  (\mathcal{U}^{-1}\mathcal{F})^{\rm T}
  \mathcal{U}^{-1}\mathcal{F}
=
\mathcal{F}^{\rm T} M^{-1}\mathcal{F}\, .
\end{equation}

\noindent
Other examples of O$(n,n)$ invariants built in the same way that arise in the
main text are

\begin{equation}
  \mathfrak{D}_{(-)\, \mu}K^{i}_{(-)\,\nu rho}K_{(-)\,\lambda\sigma}^{i}\, ,
  \,\,\,\,\,
  \text{and}
  \,\,\,\,\,
  P_{(-)\,\,ji}^{a}\mathfrak{D}_{(-)\, \mu}K^{j}_{(-)\,\nu\rho}K_{(+)\,\lambda\sigma}^{i}\, .
\end{equation}

On the other hand, if we have a O$(n)_{\pm}$ vector $\xi^{i}$, we can
construct a O$(n,n)$ vector using the Vielbein

\begin{equation}
  \mathcal{U}^{\pm}_{i}\xi^{i}
  =
  \tfrac{1}{\sqrt{2}} \left(
    \begin{array}{c}
     -e^{m}{}_{i}\xi^{i} \\ \\ \pm e_{mi}\xi^{i}-B_{mn}e^{n}{}_{i}\xi^{i} \\ 
    \end{array}
    \right)\, ,
\end{equation}

\noindent
and, then, the O$(n,n)$ invariants

\begin{subequations}
  \begin{align}
\left(\mathcal{U}^{\pm}_{i}\xi^{i} \right)^{T} \Omega\,\,
\mathcal{U}^{\pm}_{j}\xi^{j}
 & =
   \pm \xi^{i}\xi_{i}\, ,
    \\
    & \nonumber \\
\mathfrak{D}\left(\mathcal{U}^{\pm}_{i}\xi^{i} \right)^{T} \Omega\,\,
\mathfrak{D}\mathcal{U}^{\pm}_{j}\xi^{j}
 & =
   -\mathfrak{D}\left(e^{m}{}_{i}\xi^{i}\right)
   \mathfrak{D}\left( \pm e_{mj}\xi^{j}-B_{mn}e^{n}{}_{j}\xi^{j}\right)\, ,
  \end{align}
\end{subequations}

\noindent
the first of which is trivial. The second occurs in the main text with
$\xi_{i}=\varphi^{A}{}_{i}$ transforming under O$(n)_{(+)}$. With this
assignment,

\begin{equation}
\varphi^{A}{}_{i}K_{(+)}^{i}\, ,  
\end{equation}

\noindent
is also O$(n,n)$-invariant.

%%%%%%%%%%%%%%%%%%%%%%%%%%%%%%%%%%%%%%%%%%%%%%%%%%%%%%%%%%%%%%%%%%%%%%
%%%%%%%%%%%%%%%%%%%%%%%%%%%%%%%%%%%%%%%%%%%%%%%%%%%%%%%%%%%%%%%%%%%%%%
%%%%%%%%%%%%%%%%%%%%%%%%%%%%%%%%%%%%%%%%%%%%%%%%%%%%%%%%%%%%%%%%%%%%%%
%%%%%%%%%%%%%%%%%%%%%%%%%%%%%%%%%%%%%%%%%%%%%%%%%%%%%%%%%%%%%%%%%%%%%%


\begin{thebibliography}{99}
%%%%%%%%%%%%%%%%%%%%%%%%%%%%%%%%%%%%%%%%%%%%%%%%%%%%%%%%%%%%%%%%%%%%%%
%%%%%%%%%%%%%%%%%%%%%%%%%%%%%%%%%%%%%%%%%%%%%%%%%%%%%%%%%%%%%%%%%%%%%%
%%%%%%%%%%%%%%%%%%%%%%%%%%%%%%%%%%%%%%%%%%%%%%%%%%%%%%%%%%%%%%%%%%%%%%
%%%%%%%%%%%%%%%%%%%%%%%%%%%%%%%%%%%%%%%%%%%%%%%%%%%%%%%%%%%%%%%%%%%%%%

%\cite{Elgood:2020xwu}
\bibitem{Elgood:2020xwu}
Z.~Elgood and T.~Ort\'{\i}n,
``T duality and Wald entropy formula in the
Heterotic Superstring effective action at first order in $\alpha'$,''
JHEP \textbf{09} (2020), 026
\doi{10.1007/JHEP09(2020)026}
[\arxiv{2005.11272} [hep-th]].

% \cite{Buscher:1987sk}
\bibitem{Buscher:1987sk}
T.~H.~Buscher,
``A Symmetry of the String Background Field Equations,''
Phys.\ Lett.\ B {\bf 194} (1987) 59.
\doi{10.1016/0370-2693(87)90769-6}
%%CITATION = doi:10.1016/0370-2693(87)90769-6;%%

%\cite{Buscher:1987qj}
\bibitem{Buscher:1987qj}
T.~H.~Buscher,
``Path Integral Derivation of Quantum Duality in Nonlinear Sigma Models,''
Phys.\ Lett.\ B {\bf 201} (1988) 466.
\doi{10.1016/0370-2693(88)90602-8}
%%CITATION = doi:10.1016/0370-2693(88)90602-8;%%

%\cite{Bergshoeff:1995cg}
\bibitem{Bergshoeff:1995cg}
E.~Bergshoeff, B.~Janssen and T.~Ort\'{\i}n,
``Solution generating transformations and the string effective action,''
Class.\ Quant.\ Grav.\  {\bf 13} (1996) 321.
\doi{10.1088/0264-9381/13/3/002}
[\hepth{9506156}].
%%CITATION = doi:10.1088/0264-9381/13/3/002;%%

%\cite{Serone:2005ge}
\bibitem{Serone:2005ge}
M.~Serone and M.~Trapletti,
``A Note on T-duality in heterotic string theory,''
Phys. Lett. B \textbf{637} (2006), 331-337
\doi{10.1016/j.physletb.2006.03.081}
[\hepth{0512272} [hep-th]].

%\cite{Bedoya:2014pma}
\bibitem{Bedoya:2014pma}
O.~A.~Bedoya, D.~Marqu\'es and C.~N\'u\~nez,
``Heterotic $\alpha$'-corrections in Double Field Theory,''
JHEP \textbf{12} (2014), 074
\doi{10.1007/JHEP12(2014)074}
[\arxiv{1407.0365} [hep-th]].

%\cite{Lee:1990nz}
\bibitem{Lee:1990nz}
J.~Lee and R.~M.~Wald,
``Local symmetries and constraints,''
J. Math. Phys. \textbf{31} (1990), 725-743
\doi{10.1063/1.528801}

%\cite{Wald:1993nt}
\bibitem{Wald:1993nt}
R.~M.~Wald,
``Black hole entropy is the Noether charge,''
Phys.\ Rev.\ D {\bf 48} (1993) no.8,  R3427.
\doi{10.1103/PhysRevD.48.R3427}
[\grqc{9307038}].
%%CITATION = doi:10.1103/PhysRevD.48.R3427;%%

%\cite{Iyer:1994ys}
\bibitem{Iyer:1994ys}
V.~Iyer and R.~M.~Wald,
``Some properties of Noether charge and a proposal for dynamical black hole entropy,''
Phys.\ Rev.\ D {\bf 50} (1994) 846.
\doi{10.1103/PhysRevD.50.846}
[\grqc{9403028}].
%%CITATION = doi:10.1103/PhysRevD.50.846;%%

%\cite{Kraus:2005zm}
\bibitem{Kraus:2005zm}
P.~Kraus and F.~Larsen,
``Holographic gravitational anomalies,''
JHEP \textbf{01} (2006), 022
\doi{10.1088/1126-6708/2006/01/022}
[arXiv:hep-th/0508218 [hep-th]].

%\cite{Cano:2019ycn}
\bibitem{Cano:2019ycn}
P.~A.~Cano, S.~Chimento, R.~Linares, T.~Ort\'{\i}n and P.~F.~Ram\'{\i}rez,
``$\alpha'$ corrections of Reissner-Nordstr\"om black holes,''
JHEP \textbf{02} (2020), 031
\doi{10.1007/JHEP02(2020)031}
[\arxiv{1910.14324} [hep-th]].

%\cite{Maharana:1992my}
\bibitem{Maharana:1992my}
J.~Maharana and J.~H.~Schwarz,
``Noncompact symmetries in string theory,''
Nucl. Phys. B \textbf{390} (1993), 3-32
\doi{10.1016/0550-3213(93)90387-5}
[\hepth{9207016} [hep-th]].

%\cite{Maldacena:1996gb}
\bibitem{Maldacena:1996gb}
J.~M.~Maldacena and A.~Strominger,
``Statistical entropy of four-dimensional extremal black holes,''
Phys. Rev. Lett. \textbf{77} (1996), 428-429
\doi{10.1103/PhysRevLett.77.428}
[\hepth{9603060} [hep-th]].

%\cite{Johnson:1996ga}
\bibitem{Johnson:1996ga}
C.~V.~Johnson, R.~R.~Khuri and R.~C.~Myers,
``Entropy of 4-D extremal black holes,''
Phys. Lett. B \textbf{378} (1996), 78-86
\doi{10.1016/0370-2693(96)00383-8}
[\hepth{9603061} [hep-th]].

%\cite{Metsaev:1987zx}
\bibitem{Metsaev:1987zx}
R.~Metsaev and A.~A.~Tseytlin,
``Order alpha-prime (Two Loop) Equivalence of the String Equations
of Motion and the Sigma Model Weyl Invariance Conditions: Dependence
on the Dilaton and the Antisymmetric Tensor,''
Nucl. Phys. B \textbf{293} (1987), 385-419
\doi{10.1016/0550-3213(87)90077-0}

%\cite{Meissner:1996sa}
\bibitem{Meissner:1996sa}
K.~A.~Meissner,
``Symmetries of higher order string gravity actions,''
Phys. Lett. B \textbf{392} (1997), 298-304
\doi{10.1016/S0370-2693(96)01556-0}
[\hepth{9610131} [hep-th]].

%\cite{Kaloper:1997ux}
\bibitem{Kaloper:1997ux}
N.~Kaloper and K.~A.~Meissner,
``Duality beyond the first loop,''
Phys.\ Rev.\ D {\bf 56} (1997) 7940
\doi{10.1103/PhysRevD.56.7940}
[\hepth{9705193}].
%% CITATION = doi:10.1103/PhysRevD.56.7940;%%

%\cite{Hohm:2014eba}
\bibitem{Hohm:2014eba}
O.~Hohm and B.~Zwiebach,
``Green-Schwarz mechanism and $\alpha'$-deformed Courant brackets,''
JHEP {\bf 1501} (2015) 012.
\doi{10.1007/JHEP01(2015)012}
[\arxiv{1407.0708} [hep-th]].
%%CITATION = doi:10.1007/JHEP01(2015)012;%%

%\cite{Marques:2015vua}
\bibitem{Marques:2015vua}
D.~Marqu\'es and C.~A.~N\'u\~nez,
``T-duality and $\alpha'$-corrections,''
JHEP {\bf 1510} (2015) 084.
\doi{10.1007/JHEP10(2015)084}
[\arxiv{1507.00652} [hep-th]].
%%CITATION = doi:10.1007/JHEP10(2015)084;%%

%\cite{Baron:2017dvb}
\bibitem{Baron:2017dvb}
W.~H.~Baron, J.~J.~Fern\'andez-Melgarejo, D.~Marqu\'es and C.~N\'u\~nez,
``The Odd story of $\alpha'$-corrections,''
JHEP \textbf{04} (2017), 078
\doi{10.1007/JHEP04(2017)078}
[\arxiv{1702.05489} [hep-th]].

%\cite{Eloy:2019hnl}
\bibitem{Eloy:2019hnl}
C.~Eloy, O.~Hohm and H.~Samtleben,
``Green-Schwarz Mechanism for String Dualities,''
Phys. Rev. Lett. \textbf{124} (2020) no.9, 091601
\doi{10.1103/PhysRevLett.124.091601}
[\arxiv{1912.01700} [hep-th]].

%\cite{Eloy:2020dko}
\bibitem{Eloy:2020dko}
C.~Eloy, O.~Hohm and H.~Samtleben,
``Duality Invariance and Higher Derivatives,''
[\arxiv{2004.13140} [hep-th]].

%\cite{Bergshoeff:1989de}
\bibitem{Bergshoeff:1989de}
E.~A.~Bergshoeff and M.~de Roo,
``The Quartic Effective Action of the Heterotic String and Supersymmetry,''
Nucl.\ Phys.\ B {\bf 328} (1989) 439.
\doi{10.1016/0550-3213(89)90336-2}
%%CITATION = doi:10.1016/0550-3213(89)90336-2;%%

%\cite{Bergshoeff:1988nn}
\bibitem{Bergshoeff:1988nn}
E.~Bergshoeff and M.~de Roo,
``Supersymmetric Chern-simons Terms in Ten-dimensions,''
Phys.\ Lett.\ B {\bf 218} (1989) 210.
\doi{10.1016/0370-2693(89)91420-2}
%%CITATION = doi:10.1016/0370-2693(89)91420-2;%%

%\cite{Ortin:2015hya}
\bibitem{Ortin:2015hya}
T.~Ort\'{\i}n,
``Gravity and Strings'', 2nd edition, 
Cambridge University Press, 2015.
%%CITATION = INSPIRE-1383727;%%

%\cite{Fontanella:2019avn}
\bibitem{Fontanella:2019avn}
A.~Fontanella and T.~Ort\'{\i}n,
``On the supersymmetric solutions of the Heterotic
Superstring effective action,''
JHEP \textbf{06} (2020), 106
\doi{10.1007/JHEP06(2020)106}
[\arxiv{1910.08496} [hep-th]].

%\cite{Scherk:1979zr}
\bibitem{Scherk:1979zr}
J.~Scherk and J.~H.~Schwarz,
``How to Get Masses from Extra Dimensions,''
Nucl.\ Phys.\ B {\bf 153} (1979) 61.
\doi{10.1016/0550-3213(79)90592-3}
%% CITATION = doi:10.1016/0550-3213(79)90592-3;%%

%\cite{Chapline:1982ww}
\bibitem{Chapline:1982ww}
G.~F.~Chapline and N.~S.~Manton,
``Unification of Yang-Mills Theory and
Supergravity in Ten-Dimensions,''
Phys. Lett. B \textbf{120} (1983), 105-109
\doi{10.1016/0370-2693(83)90633-0}

%\cite{Chamseddine:1980cp}
\bibitem{Chamseddine:1980cp}
A.~H.~Chamseddine,
``N=4 Supergravity Coupled to N=4 Matter,''
Nucl. Phys. B \textbf{185} (1981), 403
\doi{10.1016/0550-3213(81)90326-6}

%\cite{Nicolai:1980td}
\bibitem{Nicolai:1980td}
H.~Nicolai and P.~Townsend,
``N=3 Supersymmetry Multiplets with Vanishing Trace Anomaly:
Building Blocks of the N>3 Supergravities,''
Phys. Lett. B \textbf{98} (1981), 257-260
\doi{10.1016/0370-2693(81)90009-5}

%\cite{Bergshoeff:1981um}
\bibitem{Bergshoeff:1981um}
E.~Bergshoeff, M.~de Roo, B.~de Wit and P.~van Nieuwenhuizen,
``Ten-Dimensional Maxwell-Einstein Supergravity, Its Currents,
and the Issue of Its Auxiliary Fields,''
Nucl. Phys. B \textbf{195} (1982), 97-136
\doi{10.1016/0550-3213(82)90050-5}

%\cite{Kaplan:1996ev}
\bibitem{Kaplan:1996ev}
D.~M.~Kaplan, D.~A.~Lowe, J.~M.~Maldacena and A.~Strominger,
``Microscopic entropy of N=2 extremal black holes,''
Phys. Rev. D \textbf{55} (1997), 4898-4902
\doi{10.1103/PhysRevD.55.4898}
[\hepth{9609204} [hep-th]].

%\cite{Cvetic:1995uj}
\bibitem{Cvetic:1995uj}
M.~Cvetic and D.~Youm,
``Dyonic BPS saturated black holes of heterotic string on a six torus,''
Phys. Rev. D \textbf{53} (1996), 584-588
\doi{10.1103/PhysRevD.53.R584}
[\hepth{9507090} [hep-th]].
 
%\cite{Cvetic:1995yq}
\bibitem{Cvetic:1995yq}
M.~Cveti\v{c} and A.~A.~Tseytlin,
``General class of BPS saturated dyonic black holes
as exact superstring solutions,''
Phys. Lett. B \textbf{366} (1996), 95-103
\doi{10.1016/0370-2693(95)01390-3}
[\hepth{9510097} [hep-th]].

%\cite{Cvetic:1995bj}
\bibitem{Cvetic:1995bj}
M.~Cveti\v{c} and A.~A.~Tseytlin,
``Solitonic strings and BPS saturated dyonic black holes,''
Phys. Rev. D \textbf{53} (1996), 5619-5633
\doi{10.1103/PhysRevD.53.5619}
[\hepth{9512031} [hep-th]].

%\cite{Rahmfeld:1995fm}
\bibitem{Rahmfeld:1995fm}
J.~Rahmfeld,
``Extremal black holes as bound states,''
Phys. Lett. B \textbf{372} (1996), 198-203
\doi{10.1016/0370-2693(96)00063-9}
[\hepth{9512089} [hep-th]].

%\cite{Duff:1995sm}
\bibitem{Duff:1995sm}
M.~Duff, J.~T.~Liu and J.~Rahmfeld,
``Four-dimensional string-string-string triality,''
Nucl. Phys. B \textbf{459} (1996), 125-159
\doi{10.1016/0550-3213(95)00555-2}
[\hepth{9508094} [hep-th]].

%\cite{Chimento:2018kop}
\bibitem{Chimento:2018kop}
S.~Chimento, P.~Meessen, T.~Ort\'{\i}n, P.~F.~Ram\'{\i}rez
and A.~Ruip\'erez,
``On a family of $\alpha'$-corrected solutions
of the Heterotic Superstring effective action,''
JHEP \textbf{07} (2018), 080
\doi{10.1007/JHEP07(2018)080}
[\arxiv{1803.04463} [hep-th]].

%\cite{Cano:2018brq}
\bibitem{Cano:2018brq}
P.~A.~Cano, S.~Chimento, P.~Meessen, T.~Ort\'{\i}n,
P.~F.~Ram\'{\i}rez and A.~Ruip\'erez,
  ``Beyond the near-horizon limit: Stringy
  corrections to Heterotic Black Holes,''
JHEP {\bf 1902} (2019) 192.
\doi{10.1007/JHEP02(2019)192}
[\arxiv{1808.03651} [hep-th]].
%% CITATION = doi:10.1007/JHEP02(2019)192;%%

%\cite{Behrndt:1998eq}
\bibitem{Behrndt:1998eq}
  K.~Behrndt, G.~Lopes Cardoso, B.~de Wit, D.~L\"ust,
  T.~Mohaupt and W.~A.~Sabra,
  ``Higher order black hole solutions in N=2 supergravity
  and Calabi-Yau string backgrounds,''
Phys. Lett. B \textbf{429} (1998), 289-296
\doi{10.1016/S0370-2693(98)00413-4}
[\hepth{9801081} [hep-th]].

%\cite{LopesCardoso:1998tkj}
\bibitem{LopesCardoso:1998tkj}
G.~Lopes Cardoso, B.~de Wit and T.~Mohaupt,
``Corrections to macroscopic supersymmetric black hole entropy,''
Phys. Lett. B \textbf{451} (1999), 309-316
\doi{10.1016/S0370-2693(99)00227-0}
[\hepth{9812082} [hep-th]].

%\cite{LopesCardoso:1999cv}
\bibitem{LopesCardoso:1999cv}
G.~Lopes Cardoso, B.~de Wit and T.~Mohaupt,
``Deviations from the area law for supersymmetric black holes,''
Fortsch. Phys. \textbf{48} (2000), 49-64
[\hepth{9904005} [hep-th]].

%\cite{LopesCardoso:1999fsj}
\bibitem{LopesCardoso:1999fsj}
G.~Lopes Cardoso, B.~de Wit and T.~Mohaupt,
``Macroscopic entropy formulae and nonholomorphic
corrections for supersymmetric black holes,''
Nucl. Phys. B \textbf{567} (2000), 87-110
\doi{10.1016/S0550-3213(99)00560-X}
[\hepth{9906094} [hep-th]].

%\cite{Sahoo:2006pm}
\bibitem{Sahoo:2006pm}
B.~Sahoo and A.~Sen,
``$\alpha'$ corrections to extremal dyonic black holes
in heterotic string theory,''
JHEP \textbf{01} (2007), 010
\doi{10.1088/1126-6708/2007/01/010}
[\hepth{0608182} [hep-th]].

%\cite{Prester:2008iu}
\bibitem{Prester:2008iu}
P.~Dominis Prester and T.~Terzic,
``$\alpha'$-exact entropies for BPS and non-BPS extremal
dyonic black holes in heterotic string theory from
ten-dimensional supersymmetry,''
JHEP \textbf{12} (2008), 088
\doi{10.1088/1126-6708/2008/12/088}
[\arxiv{0809.4954} [hep-th]].

%\cite{Cano:2018qev}
\bibitem{Cano:2018qev}
P.~A.~Cano, P.~Meessen, T.~Ort\'{\i}n and P.~F.~Ram\'{\i}rez,
``$\alpha'$-corrected black holes in String Theory,''
JHEP {\bf 1805} (2018) 110.
\doi{10.1007/JHEP05(2018)110}
[\arxiv{1803.01919} [hep-th]].
%%CITATION = doi:10.1007/JHEP05(2018)110;%%

%\cite{Faedo:2019xii}
\bibitem{Faedo:2019xii}
F.~Faedo and P.~F.~Ram\'{\i}rez,
``Exact charges from heterotic black holes,''
JHEP \textbf{10} (2019), 033
\doi{10.1007/JHEP10(2019)033}
[\arxiv{1906.12287} [hep-th]].

%\cite{Cano:2018aod}
\bibitem{Cano:2018aod}
P.~A.~Cano, S.~Chimento, T.~Ort\'{\i}n and A.~Ruip\'erez,
``Regular Stringy Black Holes?,''
Phys. Rev. D \textbf{99} (2019) no.4, 046014
\doi{10.1103/PhysRevD.99.046014}
[\arxiv{1806.08377} [hep-th]].

%\cite{Kutasov:1998zh}
\bibitem{Kutasov:1998zh}
D.~Kutasov, F.~Larsen and R.~G.~Leigh,
``String theory in magnetic monopole backgrounds,''
Nucl.\ Phys.\ B {\bf 550} (1999) 183
\doi{10.1016/S0550-3213(99)00144-3}
[\hepth{9812027}].
%%CITATION = doi:10.1016/S0550-3213(99)00144-3;%%

%\cite{Ortin:1996bz}
\bibitem{Ortin:1996bz}
T.~Ort\'{\i}n,
``Extremality versus supersymmetry in stringy black holes,''
Phys. Lett. B \textbf{422} (1998), 93-100
\doi{10.1016/S0370-2693(98)00040-9}
[arXiv:hep-th/9612142 [hep-th]].

%\cite{Hohm:2014sxa}
\bibitem{Hohm:2014sxa}
O.~Hohm, A.~Sen and B.~Zwiebach,
``Heterotic Effective Action and Duality Symmetries Revisited,''
JHEP \textbf{02} (2015), 079
\doi{10.1007/JHEP02(2015)079}
[\arxiv{1411.5696} [hep-th]].

%\cite{Font:1990gx}
\bibitem{Font:1990gx}
A.~Font, L.~E.~Ib\'a\~nez, D.~L\"ust and F.~Quevedo,
``Strong - weak coupling duality and nonperturbative effects in string theory,''
Phys. Lett. B \textbf{249} (1990), 35-43
\doi{10.1016/0370-2693(90)90523-9}

\bibitem{kn:EMMO}
  Z.~Elgood, P.~Meessen, \'A.~Murcia and T.~Ort\'{\i}n,
  work in progress.

  %\cite{Compere:2007vx}
\bibitem{Compere:2007vx}
G.~Comp\`ere,
``Note on the First Law with p-form potentials,''
Phys. Rev. D \textbf{75} (2007), 124020
\doi{10.1103/PhysRevD.75.124020}
[\hepth{0703004} [hep-th]].

%\cite{Jacobson:2015uqa}
\bibitem{Jacobson:2015uqa}
T.~Jacobson and A.~Mohd,
``Black hole entropy and Lorentz-diffeomorphism Noether charge,''
Phys. Rev. D \textbf{92} (2015), 124010
\doi{10.1103/PhysRevD.92.124010}
[\arxiv{1507.01054} [gr-qc]].

%\cite{Prabhu:2015vua}
\bibitem{Prabhu:2015vua}
K.~Prabhu,
``The First Law of Black Hole Mechanics for Fields
with Internal Gauge Freedom,''
Class. Quant. Grav. \textbf{34} (2017) no.3, 035011
\doi{10.1088/1361-6382/aa536b}
[\arxiv{1511.00388} [gr-qc]].

%\cite{Elgood:2020svt}
\bibitem{Elgood:2020svt}
Z.~Elgood, P.~Meessen and T.~Ort\'{\i}n,
``The first law of black hole mechanics in the
Einstein-Maxwell theory revisited,''
JHEP \textbf{09} (2020), 026
\doi{10.1007/JHEP09(2020)026}
[\arxiv{2006.02792} [hep-th]].

%\cite{Aneesh:2020fcr}
\bibitem{Aneesh:2020fcr}
P.~B.~Aneesh, S.~Chakraborty, S.~J.~Hoque and A.~Virmani,
``First law of black hole mechanics with fermions,''
Class. Quant. Grav. \textbf{37} (2020) no.20, 205014
\doi{10.1088/1361-6382/aba5ab}
[\arxiv{2004.10215} [hep-th]].

%\cite{Bergshoeff:1990ax}
\bibitem{Bergshoeff:1990ax}
E.~A.~Bergshoeff and M.~de Roo,
``The string effective action in the dual
formulation of D = 10 supergravity,''
Phys. Lett. B \textbf{247} (1990), 530-536
\doi{10.1016/0370-2693(90)91896-J}

%\cite{Tachikawa:2006sz}
\bibitem{Tachikawa:2006sz}
Y.~Tachikawa,
``Black hole entropy in the presence of Chern-Simons terms,''
Class. Quant. Grav. \textbf{24} (2007), 737-744
\doi{10.1088/0264-9381/24/3/014}
[\hepth{0611141} [hep-th]].

%\cite{Edelstein:2018ewc}
\bibitem{Edelstein:2018ewc}
  J.~D.~Edelstein, K.~Sfetsos, J.~A.~Sierra-Garc\'{\i}a
  and A.~Vilar L\'opez,
  ``T~duality and high-derivative gravity theories:
  the BTZ black hole/string paradigm,''
JHEP {\bf 1806} (2018) 142.
\doi{10.1007/JHEP06(2018)142}
[\arxiv{1803.04517} [hep-th]].
%%CITATION = doi:10.1007/JHEP06(2018)142;%%

\bibitem{kn:EMPO}
Z.~Elgood, P.~Meessen, D.~Pere\~n\'{\i}guez and T.~Ort\'{\i}n,
  work in progress.

\end{thebibliography}
\end{document}